\documentclass{WileyMSP-template}
\usepackage{siunitx}
\usepackage{xcolor}
\usepackage{graphicx}
\usepackage[version=4]{mhchem}
\usepackage{amsmath}
\usepackage{comment}
\usepackage{amsfonts}
\usepackage{amssymb}
\usepackage{microtype}
\usepackage[doublespacing]{setspace}
\usepackage{threeparttable}
\usepackage{enumerate}
\usepackage{hyperref}
\usepackage{xr}
\begin{document}

\pagestyle{fancy}
\rhead{}

\title{High-Temperature and High-Speed Atomic Force Microscopy Using a qPlus Sensor in Liquid via Quadpod Scanner and Hybrid-Loop Frequency Demodulation}

\maketitle

\author{Yuto Nishiwaki}
\author{Toru Utsunomiya}
\author{Takashi Ichii*}

\begin{affiliations}
Y. Nishiwaki, Dr. T. Utsunomiya, Dr. T. Ichii\\
Department of Materials Science and Engineering, Kyoto University, Yoshida Honmachi, Sakyo, Kyoto, 606-8501, Japan.\\
Email Address: ichii.takashi.2m@kyoto-u.ac.jp
\end{affiliations}


\begin{abstract}
Atomic-resolution imaging on molten metal/solid interfaces at temperatures above \SI{200}{\degreeCelsius} was achieved using a high-temperature, high-speed atomic force microscope (AFM) equipped with a qPlus sensor. 
A tip-scanning high-speed Quadpod scanner for a large mass load of qPlus sensor (\SI{2.3}{g}) was developed to enhance thermal drift tolerance by high-speed scanning and thermal insulation from the heated specimen. This scanner has dominant resonant frequencies of \SI{7.05}{kHz} (lateral) / \SI{29.7}{kHz} (vertical) without a load. 
In addition, the Hybrid-loop frequency demodulation technique for low-resonant-frequency ($f_0$) sensors with a wider bandwidth than conventional phase-locked loop was also established, providing a demodulation bandwidth of $B_{\Delta f_\mathrm{inst}}\sim 0.26 f_0$ without exceeding the theoretical noise of the input deflection signal. Combining these techniques enabled atomic-resolution imaging on the molten Ga/\ce{PtGa_x} interface at $\sim\SI{210}{\degreeCelsius}$. The topographic images obtained at $\sim\SI{210}{\degreeCelsius}$ showed a relatively low-symmetry surface with an oblique lattice with a superstructure, which differed from the primitive rectangular lattice observed in the non-heated sample left at room temperature for \SI{96}{h}. This demonstrates that the developed high-temperature, high-speed AFM techniques for qPlus sensors enable visualization of non-aqueous liquid/solid interfaces above \SI{200}{\degreeCelsius} at atomic resolution, which has various potential applications, such as injection modeling, soldering, and the fabrication of liquid-metal-based catalysts.
\end{abstract}

\section{Introduction}

Dynamic-mode atomic force microscopy (AFM) in liquids\cite{Fukuma2005,Hansma1994,Putman1994} is a powerful tool for visualizing interfacial structure at molecular and atomic resolution on various liquid/solid interfaces. Especially, frequency-modulation (FM-) AFM\cite{Albrecht1991} using Si microcantilevers in liquid\cite{Fukuma2005,Fukuma2005a,Fukuma2005b} enabled high-resolution imaging of interfaces between various aqueous solutions/organic solvents\cite{Honda2019} and solid surfaces. 
In the typical in-liquid AFM setup, the entire Si microcantilever is immersed in the liquid, and the tip-sample interaction is measured using optical displacement detection techniques such as optical beam deflection\cite{Alexander1989,Fukuma2005b} and interferometry\cite{Martin1987,Rasool2010}. Therefore, it is inherently not suitable for highly viscous or opaque liquids because highly viscous liquids decrease the resonance quality factor ($Q$) of the cantilever\cite{Fukuma2005} and increase the minimum detectable force gradient\cite{Wutscher2011,Kobayashi2009}, in addition to the fundamental requirement for the liquid's transparency to establish the optical path.

For these highly viscous or opaque liquids, quartz-tuning-fork-based setups such as qPlus sensors\cite{Wutscher2011,Giessibl2019} offer a good alternative.\cite{Ichii2012,Ichii2021}
The qPlus sensor, combined with a relatively long tip (up to $\sim\SI{3}{mm}$\cite{Purckhauer2020,Tokitoh2025}) compared to Si microcantilevers, enables immersion of only the tip apex in the liquid and placement of the QTF in air or vacuum. In this setup, high $Q$ ($>100$) can be maintained even in highly viscous ($\SI{1000}{}-\SI{10000}{mPa\ s}$) liquids\cite{Yamada2020,Nishiwaki2024}, and no optical path is required for displacement detection. These characteristics have enabled the high-resolution analysis using qPlus sensors in highly viscous ionic liquids\cite{Ichii2014,Bao2024} and silicone oils\cite{Yamada2020,Nishiwaki2024}, as well as in opaque liquids such as cell culture media\cite{Purckhauer2020} and liquid metals\cite{Ichii2021,Amano2023,Ichii2025}. Since some of these non-aqueous liquids have melting points above room temperature, extending the analysis temperature range of in-liquid AFM beyond the boiling point of water ($\SI{100}{\degreeCelsius}$) is required in the applications for various non-aqueous liquids, such as thermoplastics, ionic liquids, solders, and liquid-metal-based catalysts\cite{Rahim2022,Carl2024}.

However, the conventional AFM setups are not optimized for high-temperature operations. Temperature drift from the substrate heater causes sensitivity drift in piezoelectric scanners, which are widely made of \ce{Pb(Zr,Ti)O3} (PZT). This becomes more pronounced at high temperatures because the piezoelectric sensitivity of PZT depends on temperature, which generally becomes steeper at higher temperatures\cite{Ochi1985,Wolf2004}.  Therefore, enhanced tolerance for thermal drift is required for high-resolution imaging at high temperatures. Also, the analysis temperature is inherently limited by the Curie temperature of PZT $T_\mathrm{c}\sim \SI{230}{\degreeCelsius}$ and the corresponding maximum operating temperature $T_\mathrm{max}\sim\SI{130}{\degreeCelsius}$\cite{PA4FKW} to avoid depolarization of PZT. For these reasons, the previous reports of high-resolution AFM analysis in liquids at temperatures over \SI{100}{\degreeCelsius} are quite limited. 

Therefore, this study aimed to achieve atomic-resolution AFM imaging of nonaqueous liquid/solid interfaces at temperatures above \SI{200}{\degreeCelsius} using qPlus sensors. To address the increased thermal drift caused by sample heating, high-speed scanning is essential to minimize the scan time and drift per image frame. Although the high-speed AFM technique using Si microcantilevers is actively researched, it mainly relies on high-speed scanners for light loading of small samples\cite{Ando2002,Ando2008,Schitter2007} or cantilevers\cite{Umakoshi2020,Matsui2026}, as well as on high-resonant-frequency (resonant frequency $f_0>\SI{1}{MHz}$) cantilevers\cite{Ando2008}. Since the overall weight of qPlus sensors combined with excitation mechanics is usually heavier than that of Si cantilevers, the conventional high-speed tip scanners are not suitable. Even in the sample-scanning setup, the total weight of the sample holder, including specimen heater and electrostatic shielding to suppress crosstalk between the scanner and qPlus sensor, exceeds that of the substrates widely used in high-speed AFM with sample-scanning configurations. Therefore, establishing an alternative high-speed scanner design applicable to heavy samples and sensors is essential.

Furthermore, the practical maximum scanning speed in AFM is limited not only by the scanner bandwidth but also by the force demodulation bandwidth. For dynamic-mode AFM techniques other than FM-AFM that do not employ frequency feedback, such as amplitude\cite{Ruppert2017} or phase\cite{Miyata2013} modulation, a variety of high-speed demodulation schemes have been established. However, in FM-AFM, which requires closed-loop frequency feedback for sensor excitation, most studies employ a phase-locked loop (PLL) for frequency demodulation. \label{page:sprious1}The bandwidth of PLL-based demodulation is typically limited to $f_0/10$ to $f_0/5$\cite{Schlecker2014} to maintain adequate loop stability, which is further degraded to $<f_0/20$ with a limited signal-to-noise ratio for sensor deflection, as in the case of the qPlus sensor in small-amplitude operation.
Other approaches include quadricorrelator-based demodulators\cite{Kobayashi2004} or digital Hilbert-transform-based frequency detectors\cite{Mitani2009,Miyata2018} combined with external self-oscillation circuits, and hybrid-mode AFM that combines phase-modulation (PM-) AFM technique with the conventional FM-AFM (hybrid PM/FM-AFM\cite{Yamamoto2023}). However, these approaches primarily assume a high cantilever resonant frequency and are not designed to maximize the demodulation bandwidth for sensors with limited $f_0$, such as qPlus sensors. Therefore, for high-speed FM-AFM using low-$f_0$ ($\sim \SI{20}{kHz}$) qPlus sensors, an alternative frequency demodulation technique that maximizes the demodulation bandwidth for the limited $f_0$ is required.
 
To overcome these challenges on the high-speed AFM using a qPlus sensor, we developed an alternative high-speed tip scanner for the large mass load of the qPlus sensor and a frequency demodulation technique that achieves a wider demodulation bandwidth than conventional PLL in this study. Furthermore, we built a high-speed qPlus-sensor-based AFM instrument equipped with these features and applied it to high-resolution analysis on the molten Ga/solid interface at temperatures exceeding \SI{200}{\degreeCelsius} to evaluate its atomic-resolution capabilities in high-temperature non-aqueous liquids.

\section{Results and Discussion}

\subsection{High-speed AFM using Quadpod tip scanner}
\begin{figure}
 \centering
 \includegraphics[width=0.9\linewidth,page=1]{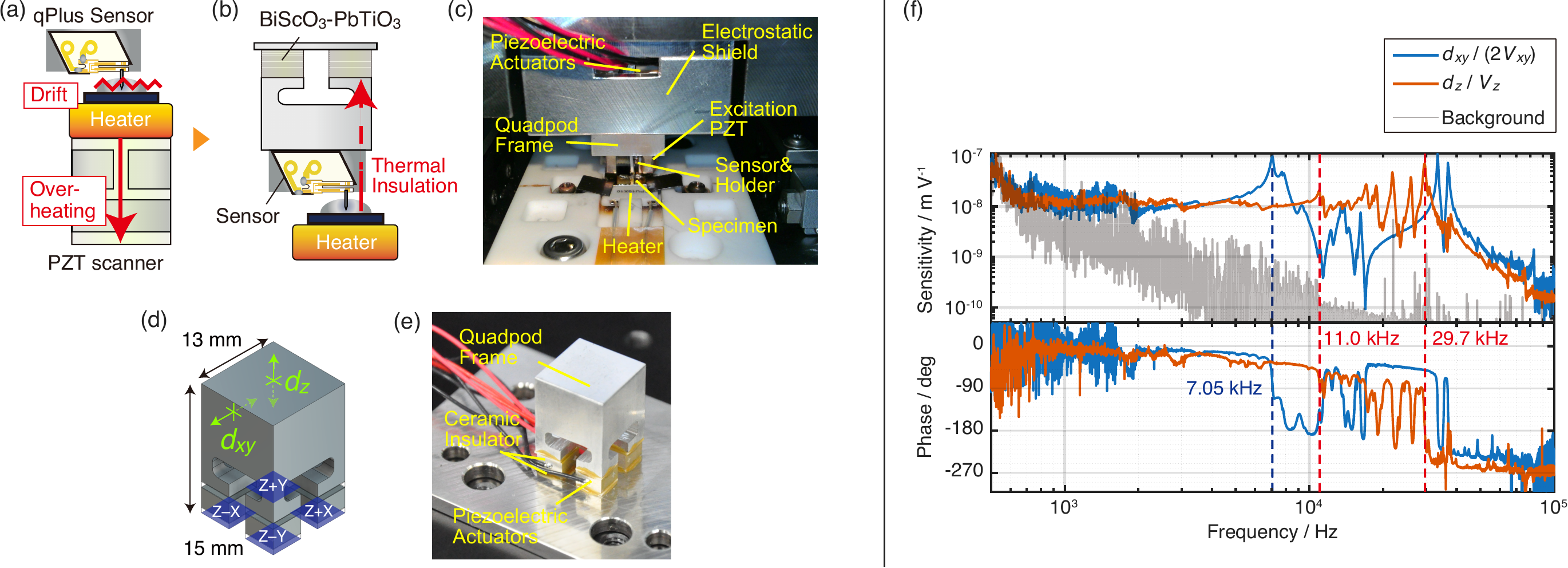}
 \caption{(a, b) Conceptual illustration of AFM investigation of heated specimen in (a) sample-scanning and (b) tip-scanning setup. (c) A photograph of the developed Quadpod-scanner-based tip-scanning AFM apparatus. (d) Conceptual diagram of the Quadpod scanner. Four actuators $\mathrm{Z\pm X}$ and $\mathrm{Z\pm Y}$ are arranged in the configuration labeled in white text. Green crosses and arrows labeled $d_{xy}$ and $d_z$ indicate the approximate displacement points and directions in the laser-Doppler velocimetry (LDV) measurement. (e) The blank Quadpod scanner for LDV measurement. (f) Bode plot of the horizontal ($d_{xy}$) and vertical ($d_z$) displacements measured at the points shown in (d), normalized to the monitored applied voltages $2V_{xy}$ and $V_z$, respectively. The gray line on the sensitivity-frequency curves indicates the equivalent floor noise in the $d_z$ direction without applied voltage. }
 \label{fig:scannerLDV}
\end{figure}

\textbf{Figure \ref{fig:scannerLDV}(a, b)} illustrates the strategy of AFM setup developed in this study. In conventional Si\\ microcantilever-based AFM, the sample-scanning setup has been widely used due to the simplicity of the optical system. 
However, since the heated sample and scanner are thermally bonded, the scanner's overheating and sensitivity drift become more problematic in high-temperature experiments (Figure \ref{fig:scannerLDV}(a)). Therefore, a tip-scanning setup (Figure \ref{fig:scannerLDV}(b)) in which the heated sample and the scanner are spatially shielded was adopted. The effectiveness of thermal insulation via the tip-scanning setup is discussed in \textbf{Supporting Information S1}. Furthermore, since the maximum operating temperature of piezoelectric materials is determined by their Curie temperature $T_c$, high-$T_c\ (=\SI{430}{\degreeCelsius})$ piezoelectric actuators made of \ce{BiScO3-PbTiO3} (BSPT)\cite{Forrester2022,Dong2023} with the operating temperature up to \SI{250}{\degreeCelsius}\cite{PA4FKYW} were employed instead of conventional \ce{Pb(Zr,Ti)O3} (PZT) actuators.

\textbf{Figure \ref{fig:scannerLDV}(c)} shows a photograph of the developed AFM apparatus. The basic structure of the AFM head, excluding the scanner, is similar to that in our previous work\cite{Tokitoh2025}, which used a tube actuator for lateral motion and a single stacked actuator for vertical motion. The entire setup was placed in a vacuum chamber at $\sim 10^1\SI{}{Pa}$ to minimize heat influx into the AFM body and thermal drift from the external environment, and to further minimize the temperature rise of the entire AFM setup during heating of the specimen. This is also discussed in Supporting Information S1. A holder with an excitation PZT for the qPlus sensor is attached to the end of the developed scanner and positioned facing the heated sample. 

The developed scanner has a ``Quadpod'' structure, as shown in \textbf{Figure \ref{fig:scannerLDV}(d)}. The four legs of the metallic quadpod frame made of A2219 aluminum alloy\cite{ASAD2013}, which is relatively suitable for high-temperature operations\cite{Polmear1988,Mondol2017}, are driven by the corresponding four stacked piezoelectric actuators. They are arranged as shown in Figure \ref{fig:scannerLDV}(d), referred to as $\mathrm{Z\pm X}$ and $\mathrm{Z\pm Y}$ hereafter, and are driven for both lateral and vertical scan with the applied voltage of $V_z \pm V_x$ and $V_z \pm V_y$, respectively.
For lateral motion, the two pairs of opposing actuators are driven in the differential mode via $\pm V_x$ and $\pm V_y$. For example, the displacement in the $d_{xy}$ direction shown in Figure \ref{fig:scannerLDV}(d) is obtained by applying an equal voltage $V_z+V_{xy}$ to two adjacent actuators of $\mathrm{Z+X}$ and $\mathrm{Z+Y}$ while applying $V_z-V_{xy}$ to the other actuators of $\mathrm{Z-X}$ and $\mathrm{Z-Y}$. For vertical displacement, which is shown as $d_{z}$ in Figure \ref{fig:scannerLDV}(d), positive common-mode voltage $V_z$ is applied in addition to $\pm V_x$ and $\pm V_y$ to all four actuators. 

To evaluate the response bandwidth of the Quadpod scanner, the frequency response under no-load conditions was measured using laser Doppler velocimetry (LDV) with a blank Quadpod scanner not equipped with a sensor holder or other components and made of PZT actuator and A7075 aluminum alloy\cite{ASAD2013}, as shown in \textbf{Figure \ref{fig:scannerLDV}(e)}. \textbf{Figure \ref{fig:scannerLDV}(f)} shows the obtained Bode plot of the horizontal ($d_{xy}$) and vertical ($d_z$) displacements, normalized to the monitored applied voltages $2V_{xy}$ and $V_z$, respectively. The approximate positions and directions of the displacement measurements are the same as those shown in Figure \ref{fig:scannerLDV}(d), and the definitions of the applied voltages $V_{xy}$ and $V_z$ during each measurement are as described above. The gray line on the sensitivity-frequency curves of the Bode plot indicates the equivalent floor noise in the $d_z$ direction without applied voltage. 

The maximum frequency where the phase shift remains within \SI{-90}{\degree}, excluding the low-frequency region where the noise floor exceeds the signal level, was \SI{7.05}{kHz} in the horizontal ($d_{xy}$) direction and \SI{11.0}{kHz} in the vertical ($d_{z}$) direction. Considering that the displacement response of a damped harmonic oscillator to an external force can be described as a second-order lag system and exhibits a \ce{-90}{\degree} phase shift at its characteristic frequency, this can be regarded as the practical maximum response bandwidth of the scanner. Also, the dominant resonant frequencies with the most pronounced sensitivity peaks were \SI{7.05}{kHz} in the horizontal direction and \SI{29.7}{kHz} in the vertical direction, which can be considered the dominant resonant frequencies for each direction.

\begin{figure}
 \centering
 \includegraphics[width=0.45\linewidth,page=2]{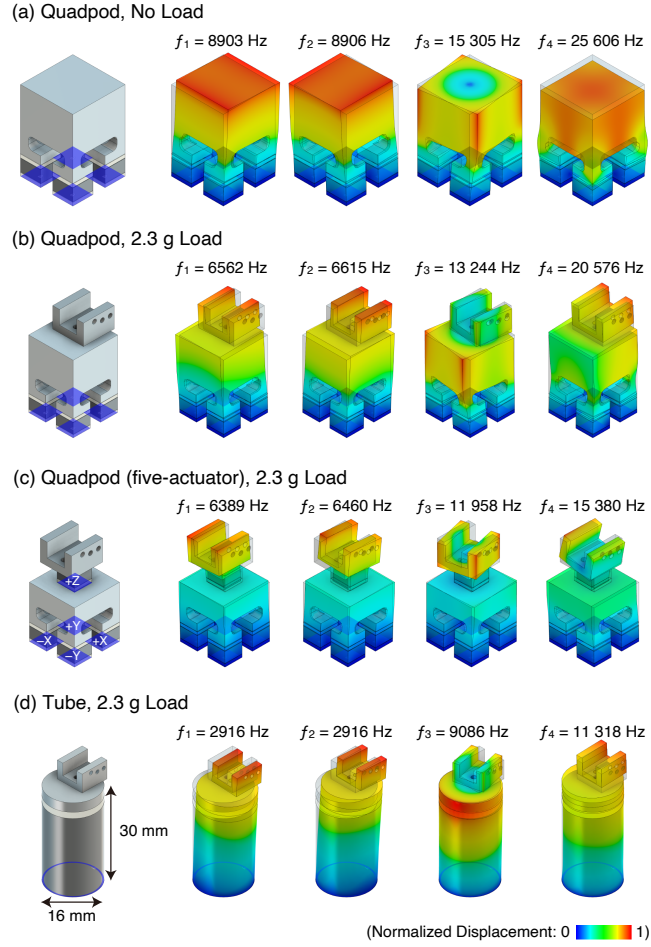}
 \caption{The normalized displacement maps for the $n$-th ($n=1,\ 2,\ 3,\ 4$) eigenmode and corresponding eigenfrequency $f_n$ of each scanner calculated in finite element method (FEM). (a) Quadpod scanner with no load, (b) Quadpod scanner with \SI{2.3}{g} load, (c) Quadpod scanner in five-actuator style with additional independent actuator for vertical scanning of \SI{2.3}{g} load. (d) Conventional tube scanner with \SI{2.3}{g} load and electrostatic shielding, modeled based on the apparatus used in our previous report\cite{Negami2013,Ichii2023}.}
 \label{fig:scannerFEM}
\end{figure}

Additionally, finite element method (FEM) simulations were performed to validate LDV measurements, characterize the load-dependent response of the Quadpod scanner, and compare it to the other types of scanners. \textbf{Figure \ref{fig:scannerFEM}} shows displacement maps for the $n$-th ($n=1,\ 2,\ 3,\ 4$) eigenmode and corresponding eigenfrequency $f_n$ of each scanner. Figure \ref{fig:scannerFEM}(a) shows the response of the Quadpod scanner without load, exhibiting resonant frequencies of $f_1\sim f_2= \SI{8.9}{kHz}$ in the horizontal direction, $f_3=\SI{15.3}{kHz}$ in the yaw direction, and $f_4=\SI{25.6}{kHz}$ in the vertical direction. 

Since FEM simulations do not account for material anisotropy, work hardening, or fixture stiffness in LDV measurements, they do not fully match the experimentally measured resonant frequencies. Nevertheless, the resonant frequency of 7.05 kHz in the $d_{xy}$ direction experimentally observed in LDV is close to that of the calculated horizontal vibration mode at $f_1\sim f_2=\SI{8.9}{kHz}$ in FEM. Also, while it is difficult to attribute all peaks detected by LDV in the $d_z$ direction, the dominant resonant frequency of \SI{29.7}{kHz} agrees with the transverse vibration mode $f_4=\SI{25.6}{kHz}$ in FEM simulations. Other minor peaks are assumed to be torsional modes and vertical vibration modes split by hardening during machining or by asymmetry due to assembly tolerances. Even considering these factors, since both $f_3$ and $f_4$ in the FEM simulation are above \SI{15}{kHz}, the \SI{11.0}{kHz} response bandwidth obtained in the LDV is sufficiently reasonable based on the FEM simulation results. Therefore, minimum resonant frequencies of at least \SI{7.05}{kHz} in the horizontal direction and at least \SI{11.0}{kHz} in the vertical direction estimated by LDV experiments are reasonable, as confirmed by FEM simulation.

Figure \ref{fig:scannerFEM}(b) shows the eigenmodes of a Quadpod scanner with a \SI{2.3}{g} stainless steel sensor holder attached as a load. Although the resonant frequencies in all four modes decreased compared to the no-load condition, the resonant frequencies were maintained at $f_1\sim f_2=\SI{6.6}{kHz}$ in the horizontal direction and $f_4=\SI{20.6}{kHz}$ in the vertical direction even with the \SI{2.3}{g} load.
For comparison, Figure \ref{fig:scannerFEM}(c) shows a different implementation of the Quadpod scanner, namely a five-actuator scanner, featuring an additional independent actuator for vertical scanning as in the common high-speed AFM scanners\cite{Schitter2007} ($+\mathrm{Z}$ Figure \ref{fig:scannerFEM}(c)) mounted between the metallic quadpod frame and the load. The horizontal resonant frequencies $f_1$ and $f_2$ were close to those of the four-actuator scanner (Figure \ref{fig:scannerFEM}(b)), whereas the resonant frequency for the torsional mode $f_3$ was notably lower than that of the four-actuator scanner. Furthermore, the fourth mode ($f_4=\SI{15.4}{kHz}$) has changed from a vertical vibration mode to a bending mode, indicating that an additional parasitic mode with a lower eigenfrequency than the vertical mode has been introduced. This occurs because the $+\mathrm{Z}$  actuator in the five-actuator configuration behaves as a necking structure with a small second moment of cross-sectional area. To increase the resonant frequency of the bending mode under these heavy loads, additional reinforcement, such as flexures\cite{Bracco2010}, is necessary when using a five-element configuration. In contrast, the four-actuator Quadpod scanner eliminates the necking structure found in the five-actuator scanner and this parasitic mode.

Similarly, Figure \ref{fig:scannerFEM}(d) shows the response of a tip scanner based on the conventional cylindrical tube scanner used in our previous report\cite{Negami2013,Ichii2023}. While the relatively large second moment of cross-sectional area of cylindrical tube prevents the introduction of parasitic bending modes below the vertical mode, the resonant frequencies were only $f_1=f_2=\SI{2.9}{kHz}$ in the horizontal direction and $f_4=\SI{11.3}{kHz}$ in the vertical direction. Compared to tube scanners, the four-actuator Quadpod scanner has a larger cross-sectional area and higher stiffness, enabling it to achieve higher resonant frequencies without being affected by additional parasitic modes, even under heavy loads, as required in applications such as qPlus-sensor-based AFM  with a tip-scanning configuration. 

\subsection{Hybrid-loop frequency demodulation for high-speed AFM using low-$f_0$ sensors}
\begin{figure}
 \centering
 \includegraphics[width=0.9\linewidth,page=3]{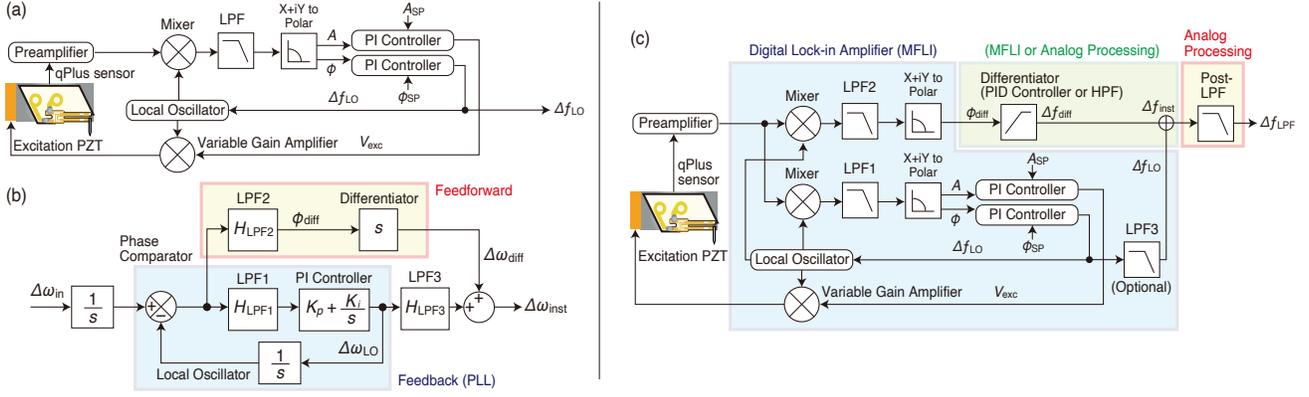}
 \caption{(a) The block diagram of the frequency demodulation and sensor excitation via phase-locked loop (PLL) using the proportional-integral (PI) controllers. (b) The block diagram of Hybrid-loop frequency demodulation. Represented as the transfer function of the angular frequency. (c) The block diagram of Hybrid-loop frequency demodulation implemented using the digital lock-in amplifier (MFLI; Zurich Instruments AG) with the additional analog processing circuit.}
 \label{fig:HybridBlock}
\end{figure}

\textbf{Figure \ref{fig:HybridBlock}(a)} shows a typical block diagram of frequency demodulation and sensor excitation using PLL on a digital lock-in amplifier, which was used in our previous study\cite{Tokitoh2025,Nishiwaki2024,Bao2024}. The frequency shift of the local oscillator $\Delta f_\mathrm{LO}$ is feedback-controlled by a proportional-integral (PI) controller so that the phase difference between the local oscillator signal, which also serves as the excitation signal, and the preamplifier output signal equals the setpoint $\phi_\mathrm{SP}\sim \SI{90}{\degree}$. This keeps the sensor excited at its resonant frequency in steady state, enabling $\Delta f_\mathrm{LO}$ to be taken as the sensor's resonant frequency shift measured by PLL. However, to ensure sufficient loop stability in this setup, the loop bandwidth $B_\mathrm{PLL}$ should be maintained at approximately $B_\mathrm{PLL}<f_0/20$ for a limited signal-to-noise ratio for sensor deflection, which limits the bandwidth of the demodulated $\Delta f_\mathrm{LO}$ signal. 

To overcome this limitation, we implemented a Hybrid-loop demodulator that combines a conventional closed-loop PLL demodulator with open-loop compensation of high-frequency residual phase difference. \textbf{Figure \ref{fig:HybridBlock}(b)} shows the block diagram of Hybrid-loop frequency demodulation representing the transfer function of the angular frequency with the linear approximation of the phase comparator.
The lower feedback-loop section of Figure \ref{fig:HybridBlock}(b) corresponds to the part of the closed-loop PLL. This part is common to the conventional PLL demodulator and provides the angular frequency shift of the local oscillator $\Delta \omega_\mathrm{LO} = 2\pi \Delta f_\mathrm{LO}$.  Note that the sinusoidal wave generated by the local oscillator is also used to mechanically drive the sensor through the excitation PZT, as in the conventional PLL-based setup. Here, the bandwidth of the low-pass filter (LPF) LPF1 $B_\mathrm{LPF1}$ is set to a sufficiently smaller value than $f_0$, typically $B_\mathrm{LPF1}\sim\SI{1}{kHz}$ for $f_0\sim\SI{20}{kHz}$ as in the conventional PLL, to ensure loop stability. This yields a $\Delta f_\mathrm{LO}$ demodulation and excitation feedback bandwidth of $B_\mathrm{PLL}\sim\SI{0.7}{kHz}$ $(=0.04 f_0)$ for the overall PLL, including the PI controller.

In contrast, the upper feedforward section branching off from the PLL loop in Figure \ref{fig:HybridBlock}(b) corresponds to an open-loop demodulator that demodulates the high-frequency residual phase difference $\phi_\mathrm{diff}$ and the corresponding residual frequency shift $\Delta f_\mathrm{diff}$. The idea of using a phase signal in addition to a frequency signal demodulated by the PLL is similar to the ``PM/FM-AFM'' setup in the previous study\cite{Yamamoto2023}. 
However, in contrast to the PM/FM-AFM setup that treats the $\phi_\mathrm{diff}$ and $\Delta f_\mathrm{LO}$ signal as the independent feedback signals for tip-sample distance control, in the Hybrid-loop frequency demodulation scheme of Figure \ref{fig:HybridBlock}(b), these signals are synthesized into a single angular frequency shift signal $\Delta \omega_\mathrm{inst}$ using the differentiator and the additional LPF (LPF3). 

To start with, for $\Delta\omega_\mathrm{LO}$ and the angular frequency of input frequency $\Delta\omega_\mathrm{in}$, the closed-loop transfer function $H_\mathrm{\Delta\omega_\mathrm{LO}}(s)=\mathcal{L}[\Delta\omega_\mathrm{LO}]/\mathcal{L}[\Delta\omega_\mathrm{in}]$ ($\mathcal{L}$: Laplace transform) is expressed as follows.
\begin{align}
 H_\mathrm{\Delta\omega_\mathrm{LO}}(s)&=\frac{1}{s}\frac{H_\mathrm{LPF1}(s)\left(K_p+\frac{K_i}{s}\right)}
   {1+H_\mathrm{LPF1}(s)\left(K_p+\frac{K_i}{s}\right)\frac{1}{s}}
\end{align}
Here, $H_\mathrm{LPF1}(s)$ is the transfer function of LPF1, and $K_p$ and $K_i$ are the proportional and integral gains of the PI controller for $\Delta\omega_\mathrm{LO}$ feedback, respectively. 
Also, for the residual angular frequency shift $\Delta \omega_\mathrm{diff}=2\pi \Delta f_\mathrm{diff}$, the transfer function $H_\mathrm{\Delta\omega_\mathrm{diff}}(s)=\mathcal{L}[\Delta\omega_\mathrm{diff}]/\mathcal{L}[\Delta\omega_\mathrm{in}]$ with closed-loop $\omega_\mathrm{LO}$ feedback is expressed using the transfer function of LPF2 $H_\mathrm{LPF2}(s)$ as follows.
\begin{align}
 H_\mathrm{\Delta\omega_\mathrm{diff}}(s)&=\frac{1}{s}\frac{sH_\mathrm{LPF2}(s)}
   {1+H_\mathrm{LPF1}(s)\left(K_p+\frac{K_i}{s}\right)\frac{1}{s}}
\end{align}
Now, the transfer function for the output signal of the whole Hybrid-loop demodulator $\Delta\omega_\mathrm{inst}=2\pi \Delta f_\mathrm{inst}$ defined in Figure \ref{fig:HybridBlock}(b) is expressed using the transfer function of LPF3 $H_\mathrm{LPF3}(s)$ as follows.
\begin{align}
 H_\mathrm{\Delta\omega_\mathrm{inst}}(s)
 &=H_\mathrm{\Delta\omega_\mathrm{diff}}(s)+H_\mathrm{LPF3}(s)H_\mathrm{\Delta\omega_\mathrm{LO}}(s)
 \\
 &=H_\mathrm{LPF2}(s)
 \frac{1+\frac{H_\mathrm{LPF3}(s)}{H_\mathrm{LPF2}(s)}H_\mathrm{LPF1}(s)\left(K_p+\frac{K_i}{s}\right)\frac{1}{s}}
   {1+H_\mathrm{LPF1}(s)\left(K_p+\frac{K_i}{s}\right)\frac{1}{s}}\label{eq:dFinst}
\end{align}
The LPF3 is intended to set to $H_\mathrm{LPF3}(s)=H_\mathrm{LPF2}(s)$, and in this situation, the whole transfer function $H_\mathrm{\Delta\omega_\mathrm{inst}}(s)$ falls into $H_\mathrm{\Delta\omega_\mathrm{inst}}(s)=H_\mathrm{LPF2}(s)$ regardless of the cleosed-loop parameters $H_\mathrm{LPF1},\ K_p$, and $K_i$. That is, the frequency demodulation bandwidth $B_{\Delta f_\mathrm{inst}}$ of $\Delta f_\mathrm{inst}$ can be freely set by LPF2 and the equivalent LPF3, within the range where the linear approximation of the phase comparator holds, regardless of the PLL loop bandwidth $B_\mathrm{PLL}$.

\label{page:sprious2}Therefore, the maximum $B_{\Delta f_\mathrm{inst}}$ is limited only by the need to be sufficiently small to separate the demodulated fundamental zero-frequency component from the mixer (complex multiplier) output from the harmonic $2f_0$ frequency component\cite{ZurichPrinciples}. Now, based on Carson's bandwidth law\cite{Carson1922}, the bandwidth of both of these components can be coarsely approximated as $2(\Delta f_\mathrm{max} +f_m)$, where $\Delta f_\mathrm{max} (\sim B_\mathrm{PLL})$ is the maximum baseband frequency shift and $f_m$ is the highest modulation signal frequency. Hence, the ultimate modulation signal frequency, where the zero-frequency component is always greater than the $2f_0$ component, is $f_m = f_0 - \Delta f_\mathrm{max}$, and this becomes the theoretical maximum bandwidth of Hybrid-loop demodulation $B_{\Delta f_\mathrm{inst}}$. This converges to $f_0$ when $\Delta f_\mathrm{max}$ is sufficiently small, and in the practical setup, the residual $2f_0$ component can be removed by applying the steep post-LPF $\Delta f_\mathrm{inst}$ at the cutoff frequency adequately less than $f_0$, which yields the post-filtered signal $\Delta f_\mathrm{LPF}$. 

This demodulation scheme can be implemented directly in an FPGA-based digital lock-in amplifier with flexible routing functions. \textbf{Figure \ref{fig:HybridBlock}(c)} shows an example implementation using a commercial digital lock-in amplifier (MFLI; Zurich Instruments AG). The differentiator was implemented using the derivative term of the lock-in amplifier's built-in proportional-integral-derivative (PID) controller, and the LPF3 and adder were implemented using the additional demodulator channel combined with the flexible routing functionality (see Experimental Section). However, even for lock-in amplifiers with lower routing flexibility or lacking additional LPFs or PID controllers, the differentiator may be replaced with a first-order analog high-pass filter (HPF) with sufficiently wide bandwidth, and LPF3 may be omitted. In this case, if the bandwidth of LPF2 is sufficiently high relative to the frequency of the signal of interest, then under the approximation $H_\mathrm{LPF2}\sim H_\mathrm{LPF3}=1$, $H_\mathrm{\Delta\omega_\mathrm{inst}}(s)$ in Equation \eqref{eq:dFinst} can be approximated as $H_\mathrm{LPF2}(s)$ as in the case where LPF3 presents.

\begin{figure}
 \centering
 \includegraphics[width=0.45\linewidth,page=4]{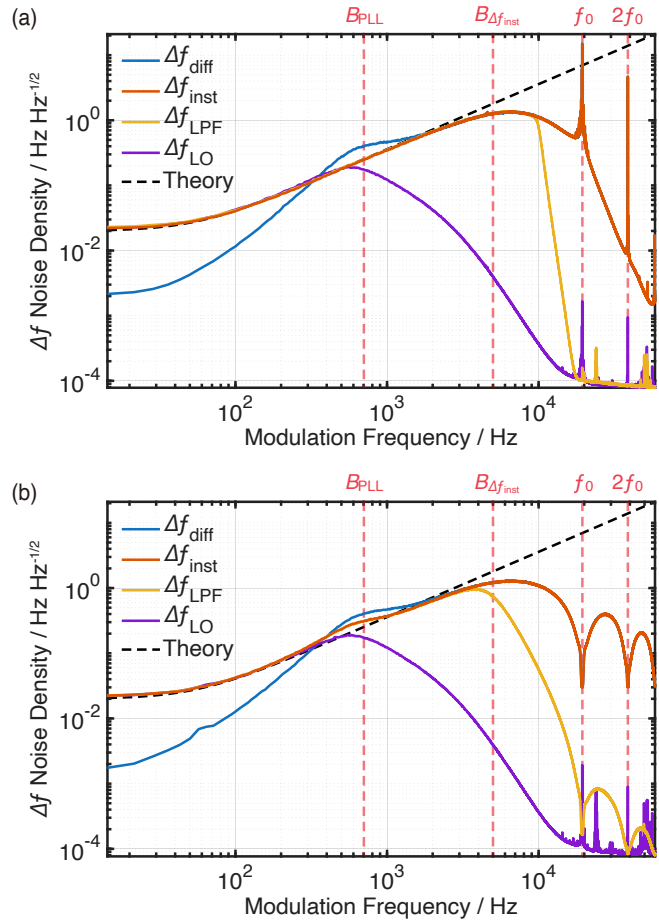}
 \caption{$\Delta f_\mathrm{diff}$, $\Delta f_\mathrm{inst}$, $\Delta f_\mathrm{LPF}$, and $\Delta f_\mathrm{LO}$ noise spectrum obtained using the setup shown in Figure \ref{fig:HybridBlock}(c) with the actual qPlus sensor in air. (a) Using cascaded exponential filters as LPF1, LPF2, and LPF3, a PID-controller-based differentiator, and an eighth-order elliptic (Cauer) filter with the passband bandwidth of 9.4 kHz as post-LPF.  (b) Using a cascaded exponential filter as LPF1, a Sinc filter as LPF2, an analog-HPF-based differentiator, a fourth-order Butterworth filter with \SI{-3}{dB} bandwidth of 5 kHz as post-LPF, and without LPF3.  The ``Theory'' curve shows the theoretical noise spectrum curve\cite{Kobayashi2009} of the input deflection signal, estimated from the thermal noise spectrum. $f_0=\SI{19395}{Hz}$, $Q=1918$, $A_\mathrm{p-p}$ (peak-to-peak) $\sim\SI{267}{pm}$.}
 \label{fig:dFspectrum}
\end{figure}

\textbf{Figure \ref{fig:dFspectrum}(a)} shows the $\Delta f_\mathrm{diff}$, $\Delta f_\mathrm{inst}$, $\Delta f_\mathrm{LPF}$, and $\Delta f_\mathrm{LO}$ noise spectrum measured using a qPlus sensor placed in air and a Hybrid-loop demodulator in Figure \ref{fig:HybridBlock}(c), implemented using the PID-controller-based differentiator and LPF3. 
These spectra were measured with LPF2 and LPF3 configured as an equivalent cascaded exponential filter with the \SI{-3}{dB} bandwidth of $B_\mathrm{LPF2}=\SI{5}{kHz}$ and an analog post-LPF configured as an eighth-order elliptic (Cauer) filter with the passband bandwidth of 9.4 kHz and the minimum stopband attenuation of \SI{106}{dB}. Also, the theoretical frequency noise spectrum\cite{Kobayashi2009} of the input deflection signal estimated from the thermal vibration spectrum of the qPlus sensor is shown in Figure \ref{fig:dFspectrum}(a). 
In the $\Delta f$ spectrum, $\Delta f_\mathrm{LO}$ follows the theoretical curve\cite{Kobayashi2009} up to the demodulation bandwidth $B_\mathrm{PLL}$, as in the conventional PLL. In contrast, $\Delta f_\mathrm{diff}$, the high-frequency residual signal of $\Delta f_\mathrm{LO}$ feedback in the PLL loop, exhibits a crossover peak around $B_\mathrm{PLL}$. However, in the sum of these signals, $\Delta f_\mathrm{inst}$, the crossover peak around $B_\mathrm{PLL}$ is completely suppressed from $\Delta f_\mathrm{diff}$. This results from $\Delta f_\mathrm{diff}$ and $\Delta f_\mathrm{LO}$ not being independent of each other but being antiphase at frequencies around $B_\mathrm{PLL}$. 

As a result, the $\Delta f_\mathrm{inst}$ curve follows the theoretical curve\cite{Kobayashi2009} below and even over $B_\mathrm{PLL}$. From this curve, the $\Delta f_\mathrm{inst}$ demodulation bandwidth, defined as the frequency of \SI{-3}{dB} attenuation from the theoretical noise spectrum curve\cite{Kobayashi2009}, was determined to be $B_{\Delta f_\mathrm{inst}}^{\SI{-3}{dB}}=\SI{5}{kHz}$ ($\sim0.26f_0$), which is equal to $B_\mathrm{LPF2}$. Also, while $\Delta f_\mathrm{inst}$ includes the narrow peaks corresponding to the deflection signal's DC noise at $f_0$ and harmonic components at $2f_0$, they are almost completely removed by the post-LPF in the $\Delta f_\mathrm{LPF}$ curve. These results demonstrate that Hybrid-loop demodulation enables wideband demodulation beyond $B_\mathrm{PLL}$ without exceeding the noise of an ideal demodulator by using the $\Delta f_\mathrm{LPF}$ signal. This represents an advantage of Hybrid-loop frequency demodulation over directly using the $\Delta f_\mathrm{diff}$ or $\phi_\mathrm{diff}$ signal with crossover peaking as an independent feedback signal, as in hybrid PM/FM-AFM\cite{Yamamoto2023}, in addition to the simplicity of parameter tuning.

Another simplified implementation that does not require a high-order analog post-LPF is to use the Sinc filter\cite{Small1994}, which is incorporated into many commercial digital lock-in amplifiers. Sinc filter is a type of digital finite impulse response (FIR) filter and has the periodic notches at integer multiples of $f_0$, which can be used for the rejection of $f_0$ and $2f_0$ peaks by using as LPF2 in Figure \ref{fig:HybridBlock}(c).
\textbf{Figure \ref{fig:dFspectrum}(b)} shows the $\Delta f_\mathrm{diff}$, $\Delta f_\mathrm{inst}$, $\Delta f_\mathrm{LPF}$, and $\Delta f_\mathrm{LO}$ noise spectrum obtained using the built-in Sinc filter in MFLI as LPF2, a first-order analog HPF with a center frequency of \SI{5.3}{kHz} as the differentiator, a fourth-order 5 kHz Butterworth filter as post-LPF, and without LPF3. 
As shown in the $\Delta f_\mathrm{inst}$ spectrum, when LPF3 is not used and a HPF with a finite center frequency is used as the differentiator, the crossover peak around the $B_\mathrm{PLL}$ is not completely eliminated in the $\Delta f_\mathrm{inst}$, but still adequately suppressed from $\Delta f_\mathrm{diff}$.
Also, in the $\Delta f_\mathrm{inst}$ curve, the $f_0$ and $2f_0$ components found in $\Delta f_\mathrm{inst}$ in Figure \ref{fig:dFspectrum}(a) were completely removed by the Sinc filter, while the residual periodic lobes away from these notch frequencies remain in the $\Delta f_\mathrm{inst}$ spectrum. However, these lobes were adequately suppressed in $\Delta f_\mathrm{LPF}$ to a level comparable to the height of the $f_0$ and 2$f_0$ peaks in $\Delta f_\mathrm{LO}$. Hereafter, in AFM experiments using Hybrid-loop demodulation, $\Delta f_\mathrm{LPF}$ in this simplified setup was employed as the feedback signal for the tip-sample distance control.

\label{page:sprious3}One of the key features of Hybrid-loop frequency demodulation is that the overall frequency demodulation $B_{\Delta f_\mathrm{inst}}$ bandwidth can be selected independently of the PLL bandwidth $B_\mathrm{PLL}$ without compromising loop stability. This is because high-frequency components above $B_\mathrm{PLL}$ are solely demodulated in the feedforward part as $\Delta f_\mathrm{diff}$ in Figure \ref{fig:dFspectrum} and are not used for the PLL feedback nor excitation, as confirmed by the block diagram in Figure \ref{fig:HybridBlock}(c). This is especially advantageous in small-amplitude operation or at high temperatures because $\Delta f_\mathrm{diff}$ is not degraded even with a limited signal-to-noise ratio in these operations, as long as $B_\mathrm{PLL}$ is sufficiently small to maintain loop stability. 

\label{page:sprious4}On the other hand, in the quantitative force spectroscopy applications such as force-distance curve measurements, $B_\mathrm{PLL}$ should not be set too small. When using high-$Q$ sensors such as the qPlus sensor ($>100$ in liquid), the phase-frequency curve is steep near the resonance frequency, resulting in a smaller $\Delta f$ error against phase error and hence a weaker coupling between the corresponding conservative and dissipative forces\cite{Sader2006}. However, in regions where the phase offset from the setpoint $\phi_\mathrm{SP}$ is nearly $\pm\SI{90}{\degree}$ (i.e., PLL is unlocked) due to insufficient PLL feedback bandwidth, the phase-frequency curve is no longer steep, and the coupling between conservative and dissipative forces cannot be ignored. Since the phase error $\phi_\mathrm{diff}$ is described as the integral of the high-frequency phase residual $\Delta f_\mathrm{diff}$, the propagation of the phase error $\phi_\mathrm{diff}$ becomes larger in the low-frequency region for the same frequency shift $\Delta f_\mathrm{diff}$. Therefore, setting $B_\mathrm{PLL}$ to a value comparable to that used in conventional FM-AFM imaging with $\Delta f_\mathrm{LO}$ is a reasonable starting point for ensuring a proper feedback bandwidth. Another best practice is to record $\phi_\mathrm{diff}$ to confirm that no noticeable phase unlocking has occurred. However, if a feedback bandwidth sufficient to prevent unlocking is unachievable, offline decoupling can be performed after observation as in hybrid PM/FM-AFM\cite{Yamamoto2023}. This would be further necessary under low-$Q$ conditions, such as Si microcantilevers in liquid, where the slope of the phase-frequency curve is small even near the resonance point and thus accepts smaller phase errors for the same frequency shift.

\subsection{AFM investigations}
As a benchmark test for atomic-resolution capabilities in non-aqueous liquids, we performed high-resolution topographic imaging on the molten \ce{Ga/MGa_x} (M = Au, Pt) interface at room temperature and at $\sim$\SI{210}{\degreeCelsius}, formed at the contact between a Ga droplet and a clean Au- or Pt-deposited mica substrate. An Au-deposited substrate was used for the room-temperature benchmark, as our previous research has confirmed that this setup exposes the \ce{AuGa2}(111) surface with a sixfold atomic arrangement on the interface.\cite{Ichii2021} \label{page:whyPt}For benchmarking at high temperatures of around \SI{210}{\degreeCelsius}, Pt-deposited substrates were used instead of Au-deposited substrates. This is because Pt has a lower self-diffusion coefficient than Au, which is advantageous for obtaining lower diffusion and alloying rates according to Darken's equation\cite{Sridhar2010}. The investigations were conducted under a vacuum of $\sim 10^1\ \mathrm{Pa}$ using the developed AFM setup shown in Figure \ref{fig:scannerLDV}(c). Prior to in-liquid AFM analysis, a rough calibration of the developed scanner was performed using a grid-patterned photomask with a known geometry, which is shown in \textbf{Supporting Information S2}. This calibration yielded maximum scan ranges of \SI{3.4}{\micro m} in the X direction, \SI{2.9}{\micro m} in the Y direction, and \SI{0.92}{\micro m} in the Z direction. 

\begin{figure}
 \centering
 \includegraphics[width=0.95\linewidth,page=5]{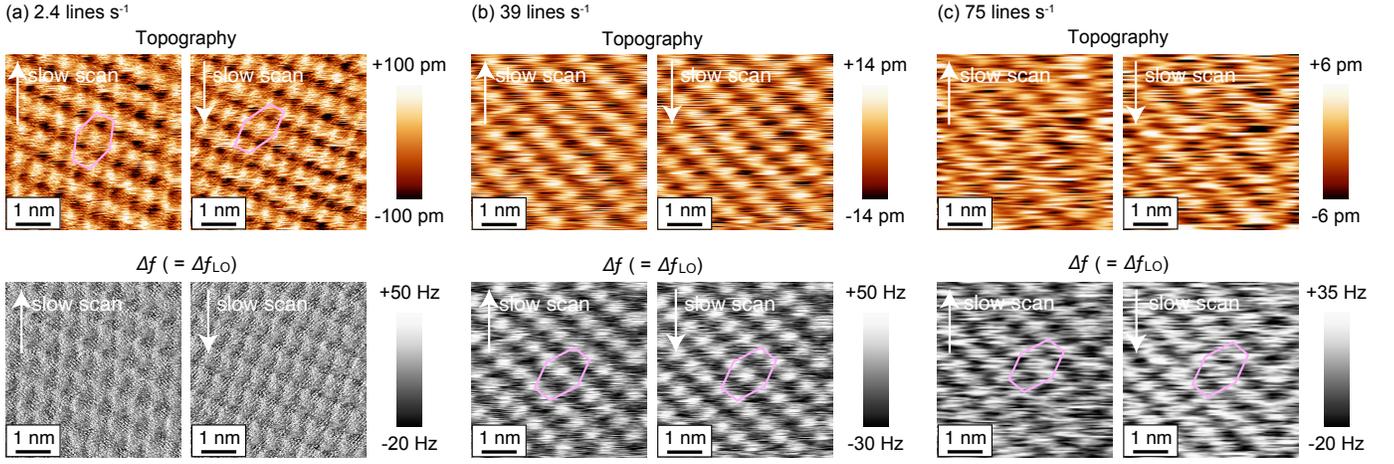}
 \caption{Topographic images and corresponding frequency shift $\Delta f$ ($=\Delta f_\mathrm{LO}$) images obtained on the Ga/\ce{AuGa2} interface in the constant-$\Delta f_\mathrm{LO}$ (conventional PLL frequency demodulation) feedback. Acquired at line scan rates of (a) \SI{2.4}{lines\ s^{-1}}, (b) \SI{39}{lines\ s^{-1}}, and (c) \SI{75}{lines\ s^{-1}}, respectively. $f_0=\SI{18163}{Hz}$, $Q=104$, $A_\mathrm{p-p}\sim\SI{208}{pm}$. Resolution (pixels$\times$lines): (a) $512\times 256$, (b) $256\times 256$, (c) $128\times 128$. The hexagonal annotations on each image indicate the positions of identical bright spots between consecutive scans determined by pattern matching, not the identical lattice of an ideal \ce{AuGa2} surface.}
 \label{fig:AuGa2}
\end{figure}

First, to evaluate the performance of the developed Quadpod scanner, we performed topographic imaging of the molten Ga/\ce{AuGa2} interface at the contact between a Ga droplet and an Au-deposited mica substrate. In this experiment, conventional PLL frequency demodulation was used instead of Hybrid-loop demodulation, and the tip-sample distance was feedback-controlled to maintain the local oscillator frequency shift $\Delta f_\mathrm{LO}$ constant (constant-$\Delta f_\mathrm{LO}$ feedback). \textbf{Figure \ref{fig:AuGa2}} shows the topographic images and the corresponding frequency shift $\Delta f (= \Delta f_\mathrm{LO})$ images of the Ga/\ce{AuGa2} interface acquired at line scan rates of 2.4, 39, and \SI{75}{lines\ s^{-1}}, respectively. The images show consecutive scans acquired in upward and downward directions for each scan rate. 
At a scan rate of \SI{2.4}{lines\ s^{-1}} (Figure \ref{fig:AuGa2}(a)), a clear contrast of a distorted hexagonal lattice pattern is visible in the topographic image, which is also faintly visible in the $\Delta f$ image. This scan rate corresponds to the imaging time of \SI{105}{s\ frame^{-1}} for 256 lines per frame, which is within the range of the typical scan rate when using the tube scanner-based AFM ($0.1-\SI{10}{lines\ s^{-1}}$\cite{Braunsmann2010}). That is, image distortion due to thermal drift cannot be ignored even at room temperature, similar to conventional AFM. Indeed, in Figure \ref{fig:AuGa2}(a), the lattice pattern annotated with hexagons in the topographic image showed a little but unignorable change between the upscan and downscan. While such relatively minor image distortions can be easily corrected by subtracting the linear drift estimated via pattern matching between consecutive scans, this becomes more difficult in high-temperature environments where the drift rate increases significantly. 

In contrast, at \SI{39}{lines\ s^{-1}} (Figure \ref{fig:AuGa2}(b)), which corresponds to the imaging time of \SI{6.6}{s\ frame^{-1}} for 256 lines per frame, both the topographic image and the $\Delta f$ image show a relatively blurred but distinct lattice contrast, whose structure agrees well between the downscan and upscan. That is, image distortion due to thermal drift has become almost negligible by increasing the scanning speed without losing atomic resolution, though resolution is constrained by the limited demodulation bandwidth of the $\Delta f_\mathrm{LO}$. The blurring of the image becomes even more pronounced at \SI{75}{lines\ s^{-1}} (Figure \ref{fig:AuGa2}(c)), which corresponds to \SI{1.7}{s\ frame^{-1}} for 128 lines per frame, making it difficult to find a clear lattice structure in the topographic image. Nevertheless, the $\Delta f$ image in Figure \ref{fig:AuGa2}(c) still exhibits a faint hexagonal lattice with the similar period as at \SI{39}{lines\ s^{-1}} (Figure \ref{fig:AuGa2}(b)), indicating that the scanner responds sufficiently fast even at such a high lateral scan rate. This demonstrates that the Quadpod scanner can potentially achieve atomic resolution at \SI{75}{Hz} and even higher scan rates, and that further improvements in scan rate are possible through $\Delta f$ demodulation bandwidth expansion.

Note that, since the piezoelectric actuators generally exhibit the sensitivity dependence on the scan range due to hysteresis and other undesirable nonlinearities\cite{Sun2014,Koops2016}, the precise calibration for nm-scale investigations should be determined apart from the \SI{}{\micro m}-scale calibration. Therefore, assuming the obtained uncalibrated atomic images in the above experiments to be \ce{AuGa2}(111) plane as confirmed in our previous work\cite{Ichii2021}, the precise calibration for nm-scale investigations was determined from these images for the following high-temperature investigations. 

\begin{figure}
 \centering
 \includegraphics[width=0.95\linewidth,page=6]{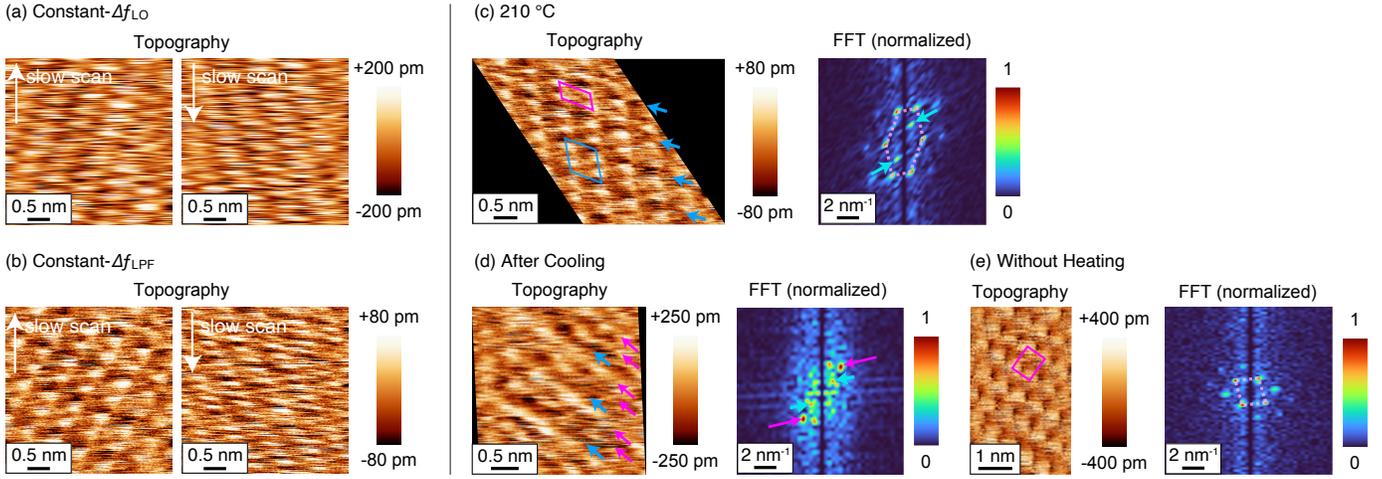}
 \caption{(a, b) Topographic images and corresponding frequency shift $\Delta f$ ($=\Delta f_\mathrm{LO},\ \Delta f_\mathrm{LPF}$) images of the consecutive scans obtained on the Ga/\ce{PtGa_x} interface at $\sim\SI{210}{\degreeCelsius}$, using (a) constant-$\Delta f_\mathrm{LO}$ feedback (conventional PLL frequency demodulation; $\Delta f_\mathrm{LO}\sim+\SI{34}{Hz}$) and (b) constant-$\Delta f_\mathrm{LPF}$ feedback (Hybrid-loop frequency demodulation; $\Delta f_\mathrm{LPF}\sim+\SI{21}{Hz}$). (c) Topographic image and its fast Fourier transform (FFT) image corresponding to the upscan (left) image of (b), after compensation for linear thermal drift and geometric tilt by pattern matching between the consecutive scans. (d) Topographic image and its FFT image obtained on the Ga/\ce{PtGa_x} interface at room temperature, \SI{50}{min} after stopping heating, following the imaging of (c). Using constant-$\Delta f_\mathrm{LPF}$ feedback ($\Delta f_\mathrm{LPF}\sim+\SI{25}{Hz}$). (e) Topographic image and its FFT image obtained on the Ga/\ce{PtGa_x} interface at room temperature, not heated after sample preparation and left in the air for \SI{96}{h}. Using constant-$\Delta f_\mathrm{LO}$ feedback ($\Delta f_\mathrm{LO}\sim+\SI{18}{Hz}$) and the conventional tube-scanner-based setup (see main text).
  (a-d) $f_0=\SI{18843}{Hz}$, $Q=400$, $A_\mathrm{p-p}\sim\SI{212}{pm}$. Scan rate: \SI{39}{lines\ s^{-1}}, imaging time: \SI{6.6}{s\ frame^{-1}}. (e) $f_0=\SI{19477}{Hz}$, $Q=210$, $A_\mathrm{p-p}\sim\SI{114}{pm}$. Scan rate: \SI{4.9}{lines\ s^{-1}}, imaging time: \SI{105}{s\ frame^{-1}}.}
 \label{fig:PtGax}
\end{figure}

Next, to evaluate the high-resolution imaging capabilities at higher temperatures, imaging at $\sim$\SI{210}{\degreeCelsius} was performed on the Ga/\ce{PtGa_x} interface at the contact between a Ga droplet and a Pt-deposited mica substrate. \textbf{Figure \ref{fig:PtGax}(a)} shows the topographic images of the consecutive scans acquired at a line scan rate of \SI{39}{lines\ s^{-1}} and an imaging time of \SI{6.6}{s\ frame^{-1}}, using the same setup as in Figure \ref{fig:AuGa2} with a constant-$\Delta f_\mathrm{LO}$ feedback. Despite the relatively high scanning speed of \SI{39}{lines\ s^{-1}}, there is a pronounced difference between consecutive upscan and downscan images due to increased thermal drift at high temperatures, in addition to the blurring similar to that shown in Figure \ref{fig:AuGa2}(b) at the same scanning speed. Although such image distortion caused by thermal drift should be corrected via pattern matching, this becomes difficult when combined with the original image's blurring.

By contrast, \textbf{Figure \ref{fig:PtGax}(b)} shows the topographic images of the consecutive scans obtained at the same scan rate, using the tip-sample distance feedback to maintain the frequency shift via Hybrid-loop demodulation $\Delta f_\mathrm{LPF}$ constant (constant-$\Delta f_\mathrm{LPF}$ feedback). The demodulator setup is the same as Figure \ref{fig:dFspectrum}(b). Similar to Figure \ref{fig:PtGax}(a), a lattice-like contrast distorted by thermal drift is visible, though it is significantly clearer than that for the constant-$\Delta f_\mathrm{LO}$ feedback image in Figure \ref{fig:PtGax}(a). This indicates that the effective spatial resolution during high-speed scanning has been enhanced by the extended frequency demodulation bandwidth provided by Hybrid-loop demodulation.

As a result, it is now possible to identify the original atomic arrangement distorted by thermal drift via pattern matching. \textbf{Figure \ref{fig:PtGax}(c)} shows the same topographic image as the upscan image of Figure \ref{fig:PtGax}(b) after compensation for linear thermal drift and geometric tilt, alongside its fast Fourier transform (FFT) image. The arrangement of bright spots in the topographic image forms an oblique lattice, as also seen as the six fundamental spots in the FFT image, highlighted by a pink parallelogram in the topographic image and a hexagon in the FFT. In addition, a series of faint dark spots, indicated by blue arrows in the topographic image, is periodically inserted in one direction between the series of bright spots. This defines a $(2\times 1)$ superlattice relative to the fundamental bright-spot lattice, annotated with a blue parallelogram in the topographic image, and introduces a pair of intense satellite peaks annotated with blue arrows in the FFT image, at half-order positions in one direction with magnitudes almost comparable to those of the fundamental spots.

Notably, this low-symmetry surface structure was not clearly seen after cooling to room temperature. \textbf{Figure \ref{fig:PtGax}(d)} shows the topographic image and the corresponding FFT image of the same specimen acquired at room temperature, \SI{50}{min} after stopping heating, following the imaging at $\sim$\SI{210}{\degreeCelsius} in Figure \ref{fig:PtGax}(c). The surface structure became more random than that observed at $\sim$\SI{210}{\degreeCelsius}, and only an indistinct stripe-like structure from the upper left to the lower right in the topographic image was observed. The bright rows indicated by the pink arrows and the periodically inserted faint dark rows indicated by the blue arrows are still faintly distinguishable. However, although the corresponding main spots indicated by the pink arrow and the faint satellite spots at half-order positions indicated by the blue arrow are still seen in the FFT image, this is not similar to that at $\sim$\SI{210}{\degreeCelsius}, where the satellite spots are almost as prominent as the main spot.

Furthermore, the non-heated samples, separately fabricated by the same procedure and left at room temperature in air for \SI{96}{h}, exhibited a surface structure different from the oblique lattice observed at $\sim$\SI{210}{\degreeCelsius}. The investigation was conducted using our conventional qPlus-sensor-based AFM setup from our previous report in air\cite{Ichii2012}, with conventional PLL frequency demodulation (constant-$\Delta f_\mathrm{LO}$ feedback) and a tube scanner. The obtained topographic image and its FFT are shown in \textbf{Figure \ref{fig:PtGax}(e)}. The topographic image showed a structure that was rather close to a primitive rectangular lattice highlighted by a pink rectangle, and the FFT showed four distinct bright spots highlighted by a pink rhombus, with no satellite spots on the low-wavenumber side. That is, two different surface structures were imaged on the Ga/\ce{PtGa_x} interface, an oblique lattice with a superstructure at $\sim$\SI{210}{\degreeCelsius} and a primitive rectangular lattice at room temperature.

If these two surface structures can be assigned on the equivalent face of the bulk crystal, the random surface structure just after cooling can be reasonably explained by assuming the temperature dependence of the stable surface structure, where the oblique lattice with superstructure is more stable at $\sim$\SI{210}{\degreeCelsius} but the primitive rectangular lattice at room temperature, respectively. However, since these intermetallic compound phases precipitate as randomly oriented polycrystals on the interface, as demonstrated by scanning electron microscopy (SEM) images in Figure S6 in \textbf{Supporting Information S3}, multiple non-equivalent surface structures can be exposed simultaneously on the different non-equivalent facets of the crystal grains. Therefore, assigning these two surface structures to the bulk crystal structure and orientation is essential for discussing the stability of each surface structure and its temperature dependence. 

Unfortunately, the phase diagram for the Pt-Ga system has not yet been conclusively determined, and even the true stable bulk structures and stoichiometries of some intermetallic phases remain unclear\cite{Tillard2011}. Indeed, in Supporting Information S3, we estimated the chemical composition of the intermetallic phase formed between a Ga droplet and a bulk Pt ribbon using energy-dispersive X-ray spectroscopy (SEM-EDS). At reaction temperatures of 30, 100, and \SI{220}{\degreeCelsius}, the estimated chemical composition was approximately \ce{Pt_{0.1}Ga_{0.9}}, where the closest match in the widely accepted phase diagram\cite{Li2006,Okamoto2007} is \ce{PtGa6}\cite{Bhan1960} with the undetermined bulk crystal structure\cite{Tillard2011}. Therefore, further characterization of the stable bulk phases and their crystal structures would enable elucidation of the relationship between bulk and the observed surface structures on the Ga/\ce{PtGa_x} interface, as well as the mechanisms of surface structure transitions. This could yield valuable insights for applications such as liquid-metal-based catalysts\cite{Rahim2022,Carl2024}, which go beyond simple AFM performance benchmarks.

\section{Conclusion}
In this study, atomic-resolution imaging on non-aqueous liquids at temperatures above \SI{200}{\degreeCelsius} was achieved by developing high-temperature, high-speed AFM techniques using a qPlus sensor. Tip-scan configuration was adopted to minimize the scanner's sensitivity drift in high-temperature operation by thermal insulation between the heated sample and the scanner, and a high-speed Quadpod scanner for a large mass load of qPlus sensor (\SI{2.3}{g}) was developed to enhance the thermal drift tolerance by high-speed scanning. 

The developed scanner has a maximum scan range of $>\SI{2.9}{\micro m}$ (lateral) / \SI{0.92}{\micro m} (vertical), as confirmed by AFM imaging of the grid-patterned photomask. Laser Doppler velocimetry demonstrated that the Quadpod scanner has dominant resonant frequencies of \SI{7.05}{kHz} (lateral) / \SI{29.7}{kHz} (vertical) and the response bandwidth with phase shift within \SI{90}{\degree} of \SI{7.05}{kHz} (lateral) / \SI{11.0}{kHz} (vertical) with no load. Finite-element-method simulations also demonstrated that these resonance frequencies, which are higher than those of conventional tube scanners, are maintained even with a mass load of \SI{2.3}{g} at \SI{6.6}{kHz} (lateral) / \SI{20.6}{kHz} (vertical). 
The high-resolution AFM imaging on the molten Ga/\ce{AuGa2} interface confirmed that Quadpod scanner can achieve atomic resolution at line scan rate of \SI{39}{line\ s^{-1}}, which could be extended to $\geq\SI{75}{line\ s^{-1}}$ when combined with higher $\Delta f$ demodulation bandwidth, and that the image distortion due to the thermal drift becomes almost negligible without losing atomic resolution when increasing the scanning speed at room temperature.

In addition, the Hybrid-loop frequency demodulation technique with a wider bandwidth than conventional PLL was also established, thereby maximizing the frequency demodulation bandwidth for the limited $f_0$ of the qPlus sensor. The $\Delta f$ noise spectrum analyses demonstrated that a demodulation bandwidth of $B_{\Delta f_\mathrm{inst}}\sim 0.26 f_0$ can be achieved without losing loop stability or exceeding the theoretical noise of the input deflection signal, theoretically up to $B_{\Delta f_\mathrm{inst}} < f_0$, which surpasses the typical PLL frequency demodulation bandwidth of $B_\mathrm{PLL}\sim 0.04 f_0$ when using a qPlus sensor with limited signal-to-noise ratio for deflection signal.

The combination of these two techniques enabled the atomic-resolution imaging on the molten Ga/\ce{PtGa_x} interface at $\sim\SI{210}{\degreeCelsius}$. The obtained topographic images showed a relatively low-symmetry surface with an oblique fundamental lattice and an associated $(2\times 1)$ superlattice, which was not clearly visible at \SI{50}{min} after cooling to room temperature. Also, this structure differed from the primitive rectangular lattice observed in the non-heated sample left at room temperature for \SI{96}{h}, suggesting that the stable structure of Ga/\ce{PtGa_x} may differ between $\sim\SI{210}{\degreeCelsius}$ and room temperature. This demonstrates that the developed high-temperature, high-speed AFM techniques for qPlus sensors enable visualization of non-aqueous liquid/solid interfaces above \SI{200}{\degreeCelsius} at atomic resolution, which has various potential applications, such as injection modeling, soldering, and the fabrication of liquid-metal-based catalysts.

\section{Experimental Section}
\threesubsection{Quadpod scanners}
Quadpod scanners were fabricated as follows. A Quadpod frame was machined from A2219 or A7075 aluminum alloy, and four \SI{0.4}{mm}-thick alumina insulators (PKFEP4; Thorlabs, Inc.) were glued to each leg using heat-resistant epoxy adhesive (EPO-TEK H74; Epoxy Technology, Inc.). Then, four $\SI{5}{mm}\times \SI{5}{mm}\times \SI{3}{mm}$ stacked piezoelectric actuators of PZT (PA4FKW; Thorlabs, Inc.) or BSPT (PA4FKYW; Thorlabs, Inc.) were glued to each leg and mounted on a metal base through four alumina insulators in the same way. 

\threesubsection{Finite Element Method (FEM)}
Finite-element-method simulation in Figure \ref{fig:scannerFEM} (and Figure S1 in the Supporting Information S1) was performed using Autodesk Fusion (Autodesk, Inc.) and its built-in materials library. Table \ref{tab:FEM-parameters} shows the thermal and mechanical properties of PZT, epoxy adhesive (EPO-TEK H74), and A2219 aluminum alloy, which is not included in the built-in library, along with the well-known values for A7075. Since the specific material properties of the PZT and BSPT actuators used in the experiments are not disclosed, the material property values of the PIC255 (PI Ceramic GmbH)\cite{PImat} used in similar commercially available stacked piezoelectric actuators\cite{PIstacked} were employed. 
Also, since the thermal and mechanical parameters of the aluminum alloys A7075 and A2219 near room temperature are quite similar, as shown in Table \ref{tab:FEM-parameters}, the built-in parameters for A7075 in Autodesk Fusion were used for both A2219 and A7075 in the FEM simulations. That is, the difference in properties between A2219 and A7075, as well as between BSPT and PZT, was assumed to be negligible in this simulation. 

\begin{table}
 \centering
  \caption{Physical Properties of PZT, EPO-TEK H74, and A2219 Alloy}
  \label{tab:FEM-parameters}
 \begin{threeparttable}
    \begin{tabular}{cccccc}
      \hline
                            &PZT\cite{PImat}\tnote{a)}              &H74\cite{H74TDS}        &A2219\cite{ASAD2013} &(A7075\cite{ASAD2013}) &(unit)\\\hline
      Thermal Conductivity &1.1                                    &1.3                     &$121-172$              &130                  &$\mathrm{W\ m^{-1}\ K^{-1}}$\\
      Young's Modulus      &52.6\tnote{b)}                         &5.93\tnote{c)}          &73                   &72                   &GPa\\
      Poisson's Ratio      &0.34                                   &0.30\cite{EpotekTechTip}&---                  &---                  &\\
      Density              &7.80                                   &2.05\tnote{d)}          &2.84                 &2.81                 &$\mathrm{g\ cm^{-3}}$\\\hline
    \end{tabular}
    \begin{tablenotes}
      \item[a)] Taken from PI255 (PI Ceramic GmbH) data.
      \item[b)] Elastic stiffness coefficient in the poling direction under a constant electric field ($C_{33}^{E}$).
      \item[c)] Storage modulus measured by dynamic mechanical analysis.
      \item[d)] Calculated from the density before mixing (Part A: $\SI{2.11}{g\ cm^{-3}}$, Part B: $\SI{1.02}{g\ cm^{-3}}$\cite{H74TDS}) and the mixing ratio by weight (100:3), ignoring volume changes.
    \end{tablenotes}
  \end{threeparttable}
\end{table}

\threesubsection{Laser-Doppler Velocimetry (LDV)}
The laser-Doppler velocimetry in Figure \ref{fig:scannerLDV} was conducted using MSA-500 Micro System Analyzer (MSA-500-TPM2-20-D-KU; Polytec GmbH), a microscope-based vibrometer. The Bode plot was obtained using an integrated frequency response analyzer (FRA) with a signal generator in MSA-500. The signal generator output was amplified using a homemade power-operational-amplifier-based voltage amplifier to drive the scanner, and the amplified signal was monitored and used as the reference signal to calculate the sensitivity and phase shift in the Bode plot. 

\threesubsection{Phase-locked Loop (PLL) and Hybrid-loop Demodulators}
The phase-locked loop (PLL) and Hybrid-loop demodulator shown in Figure \ref{fig:HybridBlock} were implemented based on the digital lock-in amplifier MFLI (Zurich Instruments AG), which incorporates four demodulator units and four PID controllers with the internal routing functionality. The proportional-integral-derivative (PID) controller-based differentiator in Figure \ref{fig:HybridBlock} was implemented by setting the integral and proportional gains of the PID controller to zero. Alternatively, in the analog-differentiator-based setup, an op-amp-based first-order high-pass filter (HPF) with a center frequency of \SI{5.3}{kHz} was used as the differentiator. Also, since the LPFs in MFLI are always placed after the mixer within each demodulator unit and cannot be used separately, a workaround was used for LPF3 in Figure \ref{fig:HybridBlock} to set the local oscillator's frequency of the corresponding mixer to zero, which effectively bypass the mixer and allows the demodulator unit to be used as a simple LPF. In addition, a fourth-order Butterworth filter and an eighth-order Elliptic filter, based on the analog filter modules SR-4BL2 and RT-8FLB2 (NF Corp.), respectively, were used as post-LPF in Figure \ref{fig:HybridBlock}(c). 

\threesubsection{AFM Setup and Investigation}
This paper employed two AFM setups: a Quadpod scanner-based setup for operation in vacuum and at variable temperatures, and a conventional tube scanner-based setup for operation in air at room temperature. The experiments in Figure 4 and 6(e) used a conventional tube-scanner-based AFM (JSPM-5200; JEOL Ltd.), which operates in ambient air and is combined with a custom-built AFM head for qPlus sensors, as introduced in our previous study\cite{Ichii2012}. The other images shown in Figure 5 and Figure 6(a-d) were acquired using the Quadpod scanner-based apparatus shown in Figure 1(c), developed in this study. In both setups, the preamplifier for the qPlus sensor, based on the design by Huber and Giessibl\cite{Huber2017}, was embedded in the AFM head, as shown in Figure S1(a) in the Supporting Information S1 for the Quadpod scanner-based setup. The frequency shift ($\Delta f$) noise spectrum in Figure \ref{fig:dFspectrum} was measured using the FFT analyzer function built into the MFLI. Also, for tip-sample distance feedback and lateral scan in topographic imaging, the Nanonis SPM Control system (SPECS Surface Nano Analysis GmbH) with high-voltage amplifier (HVA4) was used.

For the Quadpod scanner-based apparatus, the entire setup was placed in a vacuum chamber maintained at $\sim 10^1\SI{}{Pa}$ by a multi-stage roots vacuum pump (NeoDry15G; Kashiyama Industries, Ltd.). The specimen was heated using a positive-temperature-coefficient (PTC) heater (SCPU10X10; Kashima Co. Ltd.). A constant DC voltage was applied, and the temperature was monitored via its resistance, as described in Supporting Information S1.

The fabrication procedure for the qPlus sensor is the same as in our previous work\cite{Nishiwaki2025}. A \SI{32768}{Hz} quartz tuning fork (QTF; SII Crystal Technology Inc.) was mounted on an alumina substrate with the printed wiring pattern (Tec Corp.). Then, a tungsten wire with an electropolished tip, \SI{0.1}{mm} in diameter and $\sim\SI{0.8}{mm}$ in length, was glued to the free prong of the QTF using heat-resistant epoxy adhesive EPO-TEK H70E (Epoxy Technology, Inc.). 

The AFM specimen was prepared as follows. Au or Pt was vacuum-deposited (base pressure $\sim 10^{-6}~\SI{}{Pa}$ and evaporation rate $\sim \SI{0.1}{nm\ s^{-1}}$) onto a cleaved mica substrate to a thickness of $100-\SI{200}{nm}$ at room temperature, followed by the drop-coating of molten gallium (\SI{99.9999}{\percent}; Nilaco Corp.) in air at room temperature. Subsequently, AFM analysis was performed by immersing the tip apex into the molten gallium, following the same procedure as in our previous work\cite{Ichii2021,Ichii2025}.

\medskip
\textbf{Supporting Information} \par 

The following Supporting Information is available online.
\begin{enumerate}[S1.]
  \item FEM thermal analysis and the noise performance evaluation in high-temperature operation.
  \item \SI{}{\micro m}-scale calibration and maximum scan range evaluation for Quadpod scanner.
  \item SEM-EDS analyses of microstructure and chemical composition of the intermetallic compounds on the Ga/\ce{PtGa_x} interface.
\end{enumerate}

\medskip
\textbf{Acknowledgements} \par
The laser-doppler velocimetry measurement was conducted in Nanotechnology Hub (Institute for Chemical Research) in Kyoto University, supported by ``Advanced Research Infrastructure for Materials and Nanotechnology in Japan (ARIM)'' of the Ministry of Education, Culture, Sports, Science and Technology (MEXT), Proposal Number JPMX1225KT1731 and JPMX1225KT2050. This work was supported by JSPS KAKENHI Grant Number JP23K26543, JST PRESTO JPMJPR25J2, and JST SPRING Grant Number JPMJSP2110.

\textbf{Conflict of Interest Disclosure} \par
The authors have no conflicts to disclose.

\textbf{Data Availability Statement} \par
The data that support the findings of this study are available from the corresponding author upon reasonable request.

\medskip

\bibliographystyle{MSP}

\begin{figure}[p]
\textbf{Table of Contents}\\
\medskip
 \includegraphics[page=12,width=0.8\hsize]{Figures-clean-crop.pdf}
 \medskip
 \caption*{Atomic-resolution imaging on molten Ga/\ce{PtGa_x} interfaces at $\sim$\SI{210}{\degreeCelsius} was achieved with a high-temperature, high-speed atomic force microscopy (AFM) using a qPlus sensor. Using the high-speed scanner for a large mass load of qPlus sensors (Quadpod scanner) and the wide-bandwidth frequency demodulation technique for low-resonant-frequency cantilevers (Hybrid-loop demodulation), the low-symmetry surface of Ga/\ce{PtGa_x} with an oblique lattice with superstructure was imaged.}
\end{figure}

\end{document}


\pagestyle{fancy}
\rhead{}

\title{Supporting Information for ``High-Temperature and High-Speed Atomic Force Microscopy Using a qPlus Sensor in Liquid via Quadpod Scanner and Hybrid-Loop Frequency Demodulation''}

\maketitle

\author{Yuto Nishiwaki}
\author{Toru Utsunomiya}
\author{Takashi Ichii*}

\begin{affiliations}
Y. Nishiwaki, Dr. T. Utsunomiya, Dr. T. Ichii\\
Department of Materials Science and Engineering, Kyoto University, Yoshida Honmachi, Sakyo, Kyoto, 606-8501, Japan.\\
Email Address: ichii.takashi.2m@kyoto-u.ac.jp
\end{affiliations}

\section{FEM thermal analysis and the noise performance evaluation in high-temperature operation}
To evaluate the effectiveness of thermal insulation via the tip-scanning setup and vacuum insulation, a steady-state thermal analysis by finite element method (FEM) was performed. The simulation was conducted using Autodesk Fusion (Autodesk, Inc.) and its built-in materials library, along with the additional thermal property for the epoxy adhesive and PZT (BSPT) shown in Table 1 of the main text.

\begin{figure}
 \centering
 \includegraphics[width=0.7\linewidth,page=7]{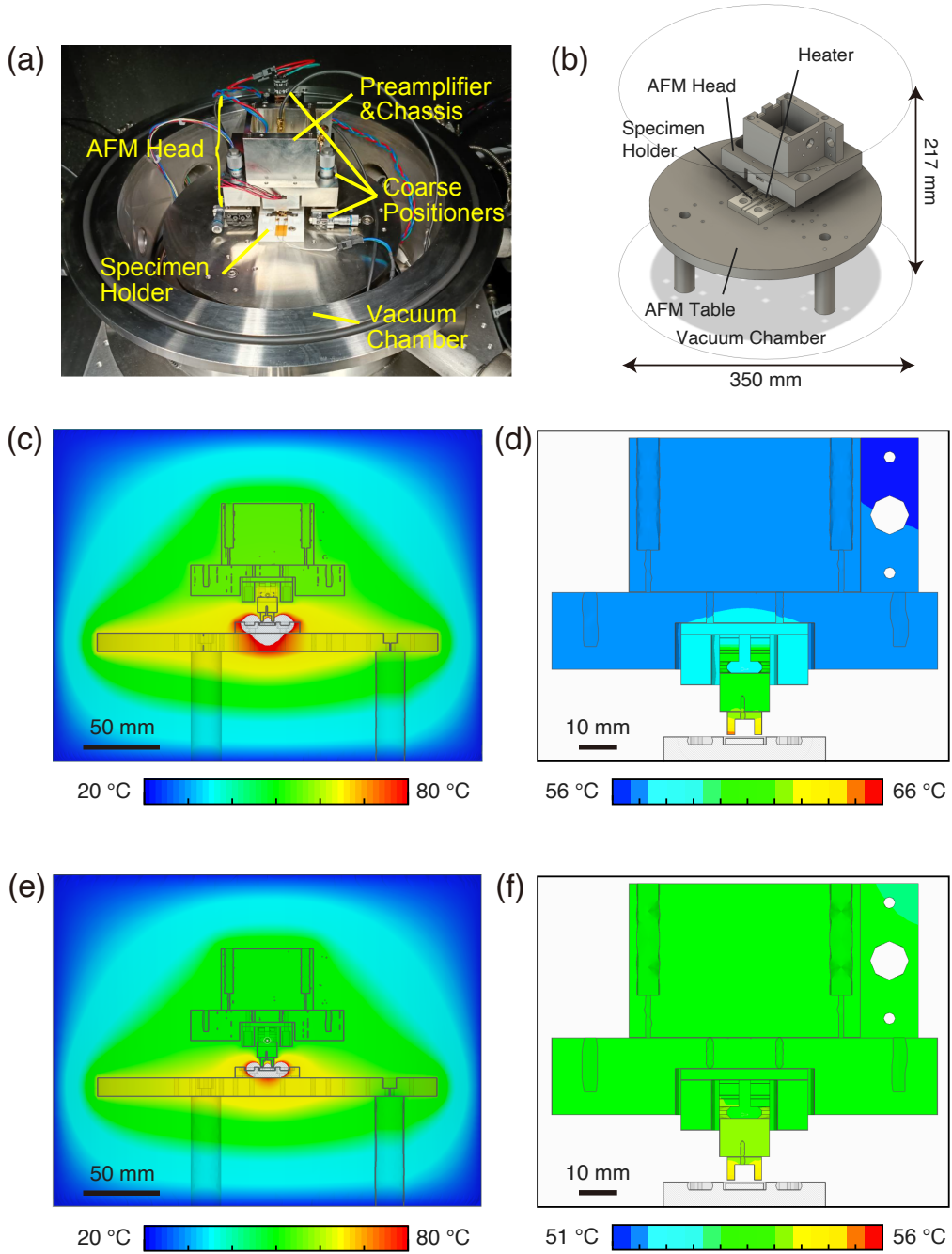} 
 \caption{(a) A photograph of the entire AFM setup placed in a vacuum chamber. (b) A snapshot of the finite element method (FEM) thermal simulation model constructed based on (a). (c, d) The cross-sectional temperature profile calculated by FEM for atmospheric pressure ($\kappa_\mathrm{atm}=\SI{2.6e-2}{W\ m^{-1}\ K^{-1}}$). (c) for the entire setup, (d) for AFM head excerpted from (c). (e, f) The cross-sectional temperature profile calculated by FEM for vacuum pressure of $\sim \SI{18}{Pa}$ ($\kappa_\mathrm{vac}=\SI{1.0e-2}{W\ m^{-1}\ K^{-1}}$). (e) for the entire setup, (f) for AFM head excerpted from (e).}
 \label{fig:ThermalFEM}
\end{figure}

\textbf{Figure \ref{fig:ThermalFEM}(a)} shows a photograph of the entire AFM setup placed in a vacuum chamber, and \textbf{Figure \ref{fig:ThermalFEM}(b)} shows a snapshot of the FEM simulation model constructed based on the setup in Figure \ref{fig:ThermalFEM}(a). In this model, heat transfer by convection and radiation was ignored, and the gas inside the vacuum chamber was approximated as a continuum with thermal conductivity of $\kappa_\mathrm{atm}=\SI{2.6e-2}{W\ m^{-1}\ K^{-1}}$ at atmospheric pressure and $\kappa_\mathrm{vac}=\SI{1.0e-2}{W\ m^{-1}\ K^{-1}}$ in a vacuum. According to Kaminski's formula\cite{Ogbonnaya2017}, the adopted value of $\kappa_\mathrm{vac}$ corresponds to a vacuum pressure of $\sim \SI{18}{Pa}$ for a characteristic length of \SI{0.8}{mm}, which is equal to the length of the qPlus sensor's tungsten tip. Also, the temperatures of the heater's top surface and the chamber wall were assumed uniform and fixed at \SI{200}{\degreeCelsius} and \SI{20}{\degreeCelsius}, respectively. 

\textbf{Figure \ref{fig:ThermalFEM}(c, d)} shows the calculated temperature profile along a vertical cross-section passing around the qPlus sensor's tip under atmospheric pressure. Figure \ref{fig:ThermalFEM}(c) represents the temperature distribution for the entire model, and Figure \ref{fig:ThermalFEM}(d) represents that for the AFM head excerpted from the profile of Figure \ref{fig:ThermalFEM}(c). As shown in Figure \ref{fig:ThermalFEM}(d), the temperatures of all components within the AFM head, including the scanner, preamplifier chassis, and qPlus sensor holder, were calculated to be within the range of \SI{56}{\degreeCelsius} to \SI{66}{\degreeCelsius}. Also, \textbf{Figure \ref{fig:ThermalFEM}(e)} shows the temperature distribution of the entire model under vacuum conditions, and \textbf{Figure \ref{fig:ThermalFEM}(f)} shows that for the AFM head, corresponding to Figure \ref{fig:ThermalFEM}(e). The temperature distribution of the entire model is similar to that under atmospheric pressure, while the AFM head temperature was slightly lower, ranging from \SI{51}{\degreeCelsius} to \SI{56}{\degreeCelsius}. Note that, given the maximum operating temperatures of PZT ($T_\mathrm{max}\sim\SI{130}{\degreeCelsius}$\cite{PA4FKW}) and BSPT ($T_\mathrm{max}\sim\SI{250}{\degreeCelsius}$\cite{PA4FKYW}), the risk of depolarization in piezoelectric actuators is negligible during AFM operation at sample temperatures around $\sim\SI{200}{\degreeCelsius}$. Therefore, the analysis temperature can be further extended, especially when using BSPT actuators.

On the other hand, as shown in Figure \ref{fig:ThermalFEM}(a), a preamplifier for the qPlus sensor\cite{Huber2017} is also mounted on the preamplifier chassis in the AFM head to minimize the wiring length and thus minimize the detector noise density $V_\mathrm{noise}$. This preamplifier is based on the OPA657 (Texas Instruments, Inc.), a field-effect transistor (FET)-input operational amplifier, whose input current noise increases exponentially with temperature due to higher input bias current.\cite{Blake2008} Additionally, the thermal noise for the feedback resistor connected to the operational amplifier in the preamplifier circuit also increases at high temperatures, proportional to the square root of temperature\cite{Blake2008}. Therefore, even below OPA657's maximum operating temperature of \SI{85}{\degreeCelsius}\cite{OPA657}, which is well above the preamplifier chassis's temperature calculated in the FEM simulation, the increase in $V_\mathrm{noise}$ due to the temperature rise should be considered.

Therefore, to evaluate the preamplifier's noise performance during high-temperature operation, thermal-vibration spectra were measured in the developed AFM apparatus at various temperatures, alongside measurements of the scanner's temperature. The scanner's temperature $T_\mathrm{Scanner}$ was measured using a thermocouple attached near the $d_{xy}$ point in Figure 1(d) of the main text. The AFM setups for thermal-vibration spectra measurements are the same as those in the AFM investigations in the main text. The AFM specimen of molten Ga/mica substrate was prepared by drop-coating of molten Ga on a cleaved bare mica substrate, and then the tungsten tip of the qPlus sensor was immersed in the Ga droplet. Thermal noise spectra were obtained as the preamplifier deflection voltage output, without sensor excitation, using the spectrum analyzer function of the MFLI digital lock-in amplifier (Zurich Instruments AG). The obtained voltage spectra $V_\mathrm{out}(f)$ were fitted by the following equation\cite{Kobayashi2009}, assuming the displacement-to-voltage sensitivity $S_V$ and the spring constant $k$ are independent of temperature ($k_\mathrm{B}$: Boltzmann constant).
\begin{align}
    V_\mathrm{out}(f)&=\sqrt{S_V^2\frac{2k_\mathrm{B}T_\mathrm{qPlus}}{\pi k Q f_0}\frac{1}{\left[1-\left(\frac{f}{f_0}\right)^2\right]^2+\left(\frac{f}{Qf_0}\right)}+V_\mathrm{noise}^2}\label{eq:thermal}
\end{align}
This yields the effective temperature for thermal vibration of the qPlus sensor $T_\mathrm{qPlus}$ and the voltage noise density of the amplifier $V_\mathrm{noise}$, in addition to the resonant frequency $f_0$ and quality factor $Q$. 
\label{page:thermalC1} Note that the estimation of $T_\mathrm{qPlus}$ has limited robustness, since it is subject to fluctuations in the effective spring constant $k$ caused by the oxide film growth of Ga and changes in the contact angle during measurement, which cannot be estimated independently from this fitting procedure.

\begin{figure}
 \centering
 \includegraphics[width=0.95\linewidth,page=8]{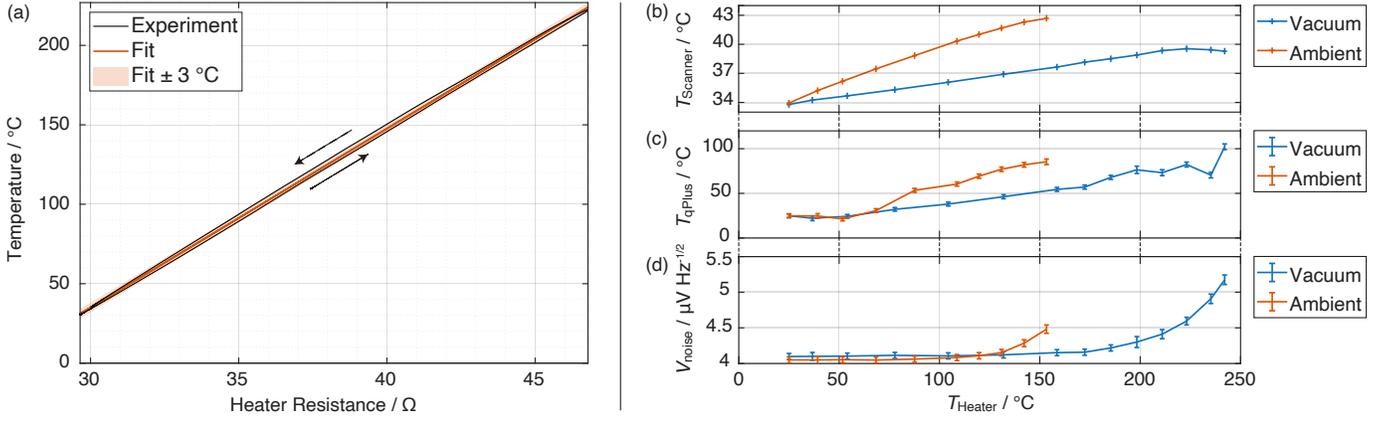}
 \caption{(a) Surface temperature-resistance curves of the PTC heater, alongside a linear fit with its $\pm\SI{3}{\degreeCelsius}$ region. The temperature was measured at the center of the heater using an external thermocouple. The arrows show the scanning direction. Total acquisition time $\sim\SI{130}{min}$. (b) Scanner's temperature $T_\mathrm{scanner}$ at each heater's temperature $T_\mathrm{Heater}$ in vacuum ($\sim\SI{10}{Pa}$) and in ambient air. $T_\mathrm{scanner}$ was measured by thermocouple, and $T_\mathrm{Heater}$ was estimated from the heater's resistance. (c-e) (c) The effective temperature for the qPlus sensor's thermal vibration $T_\mathrm{qPlus}$, (d) the voltage noise density of the amplifier $V_\mathrm{noise}$, and (e) the resonant frequency factor $Q$, simultaneously acquired with (b) in vacuum and in ambient air. Calculated from the thermal-vibration spectra obtained at each point, according to Equation \eqref{eq:thermal}. The error bar represents the \SI{95}{\percent} confidence interval in the fitting of each spectrum. $f_0=\SI{18680}{Hz},\ Q=767$ (in vacuum at room temperature, with tip immersed in Ga).}
 \label{fig:ThermalExp}
\end{figure}

In this experiment and other AFM experiments in the main text, the heater temperature $T_\mathrm{Heater}$ was estimated from its resistance, as in the previous study\cite{Broekmaat2008}. \textbf{Figure \ref{fig:ThermalExp}(a)} shows the surface temperature at the center of the heater, measured by an external thermocouple, as a function of the heater's resistance. The surface temperature was well fitted by a linear function of the resistance, with nonlinearity and hysteresis confined to $\pm\SI{3}{\degreeCelsius}$. 

\label{page:thermalB}\textbf{Figure \ref{fig:ThermalExp}(b)} shows the scanner's temperature $T_\mathrm{Scanner}$ in a vacuum and in air at each $T_\mathrm{Heater}$. These plots were acquired from the high-temperature side to the low-temperature side with measurement intervals of at least 8 min. Here, the nonlinearity in the high-temperature region is due to heat storage, as the AFM head's temperature response is extremely slow because it is thermally isolated from the AFM stage. The scanner's temperature $T_\mathrm{Scanner}$ remained below \SI{40}{\degreeCelsius} in the range where $T_\mathrm{Heater} \lesssim \SI{100}{\degreeCelsius}$ in air and $T_\mathrm{Heater} \lesssim \SI{240}{\degreeCelsius}$ in vacuum. This demonstrates that the effectiveness of thermal insulation in the tip-scan configuration is enhanced in a vacuum, which is consistent with FEM simulations. Here, the measured $T_\mathrm{Scanner}$ at $T_\mathrm{Heater} \sim \SI{200}{\degreeCelsius}$ in a vacuum was lower than that calculated in the FEM simulation of $\sim$\SI{54}{\degreeCelsius} (Figure \ref{fig:ThermalFEM}(f)). This can be attributed to the factors such as thermal radiation and the characteristic-length dependence of $\kappa_\mathrm{vac}$, which are not accounted for in the FEM simulations.

\label{page:thermalC}\textbf{Figure \ref{fig:ThermalExp}(c)} shows the estimated $T_\mathrm{qPlus}$ in a vacuum and in air at each $T_\mathrm{Heater}$. Apart from the fluctuations caused by changes in the effective spring constant $k$ mentioned earlier, $T_\mathrm{qPlus}$ increases roughly linearly with respect to $T_\mathrm{Heater}$. 
\label{page:ThermalExpCorrection}$T_\mathrm{qPlus}$ remained below $\sim\SI{200}{\degreeCelsius}$ in the whole measured temperature range of $T_\mathrm{Heater} \lesssim \SI{150}{\degreeCelsius}$ in air and $T_\mathrm{Heater} \lesssim \SI{240}{\degreeCelsius}$ in vacuum. While the maximum temperature of $T_\mathrm{qPlus}$ is even lower than the BSPT actuator's maximum operating temperature (\SI{250}{\degreeCelsius}\cite{PA4FKYW}), the measured temperature of $T_\mathrm{qPlus}=\SI{137}{\degreeCelsius}$ at $T_\mathrm{Heater}=\SI{198}{\degreeCelsius}$ in vacuum is quite higher than the FEM simulation in Figure \ref{fig:ThermalFEM}(f), and further in air, $T_\mathrm{qPlus}$ is almost equal to the $T_\mathrm{Heater}$ at $T_\mathrm{Heater}\gtrsim\SI{90}{\degreeCelsius}$. While this may be partially explained by heat transfer through the tungsten tip of the qPlus sensor or by convective heat transfer in air, the spatial weighting imposed by the thermal dissipation theorem cannot be ignored.
Specifically, the effective temperature for thermal vibrations of a cantilever with non-uniform temperature is described as an average of the temperatures at each position weighted by the mechanical energy dissipation density\cite{Fontana2020,Fontana2021,Geitner2017,Komori2018}. In this experiment, the quality factor of resonance $Q$, calculated from the thermal-vibration spectra, was $Q>2000$ in vacuum, whereas it decreased to $Q<1000$ in molten Ga at room temperature. That is, the majority of the mechanical dissipation of thermal vibration occurs at the tip apex immersed in the liquid, and the effective temperature $T_\mathrm{qPlus}$ of thermal vibrations reflects more strongly the temperature of the tip. Therefore, the average temperature of the entire qPlus sensor is expected to be maintained even below $T_\mathrm{qPlus}$.

\label{page:thermalD}\textbf{Figure \ref{fig:ThermalExp}(d)} shows the voltage noise density $V_\mathrm{noise}$ originating from the preamplifier, which was estimated simultaneously with Figure \ref{fig:ThermalExp}(c) in Equation \eqref{eq:thermal}. Unlike the nearly linear increase of $T_\mathrm{Scanner}$ or $T_\mathrm{qPlus}$ with respect to $T_\mathrm{Heater}$ in Figure \ref{fig:ThermalExp}(b, c), $V_\mathrm{noise}$ increased exponentially. The noise in an op-amp-based current amplifier can be well described as the sum of three components: thermal noise for the feedback resistor, the operational amplifier's input voltage noise $e_n$, and the input current noise $i_n$.\cite{Ramus2009} Among them, $i_n$ is the only component that can show the steep increase with respect to the temperature in this range, as the shot noise corresponding to the input bias current $i_\mathrm{B}$ of FET-input operational amplifier is proportional to $\sqrt{i_\mathrm{B}}$\cite{Blake2008}, which increases exponentially with temperature\cite{OPA657}. Thus, the exponential increase in $V_\mathrm{noise}$ above \SI{120}{\degreeCelsius} in air and above \SI{170}{\degreeCelsius} in vacuum is best explained by the increase in $i_n$ due to heating the operational amplifier. 

From the viewpoint of practical AFM investigations, the increase in $V_\mathrm{noise}$ remained within \SI{10}{\percent} over the $T_\mathrm{Heater}$ range up to \SI{211}{\degreeCelsius} in vacuum and \SI{142}{\degreeCelsius} in air, which was adequate for atomic-resolution imaging in vacuum at $\sim\SI{210}{\degreeCelsius}$, as demonstrated in Figure 6 of the main text. However, cooling the operational amplifiers or using them with lower input bias current at high temperatures would be effective for further noise reduction, potentially in higher-temperature environments.

\label{page:thermalE}On the other hand, the practical maximum temperature in the current setup is not limited by the apparatus-specific factors, such as the BSPT-based scanner's maximum operating temperature or the increase in $V_\mathrm{noise}$ at high temperatures, but rather by the qPlus sensor's heat resistance. Specifically, in the high-temperature operation where $T_\mathrm{qPlus}$ approaches or exceeds the glass transition temperature $T_g \sim 80\cite{H70ETDS}-\SI{100}{\degreeCelsius}\cite{H74TDS}$ of the heat-resistant epoxy adhesive used in the assembly of the qPlus sensor, the resonant quality factor $Q$ is degraded due to a decrease in strength, resulting in an increase in $\Delta f$ noise. 

\textbf{Figure \ref{fig:ThermalExp}(e)} shows the estimated $Q$ in a vacuum and in air at each $T_\mathrm{Heater}$. As with $T_\mathrm{qPlus}$, the $Q-T_\mathrm{Heater}$ plots show the fluctuations due to the growth of the Ga oxide film and changes in its contact angle.  
However, $Q$ roughly degrades gradually as $T_\mathrm{Heater}$ increases, and at temperatures above 200 °C in a vacuum, $Q$ falls below 400. This means that low-frequency noise in the $\Delta f$ signal increases with rising temperature not only due to increased amplifier's noise $V_\mathrm{noise}$ but also due to a decrease in $Q$, since the theoretical frequency noise spectrum of the input deflection signal is denoted by the following equation.\cite{Kobayashi2009}
\begin{align}
    N_{\Delta f}(f_m)&=\sqrt{\frac{f_0k_\mathrm{B}T_\mathrm{qPlus}}{\pi k Q A_0^2}+\frac{2n_\mathrm{ds}^2}{A_0^2}f_m^2+\frac{f_0^2n_\mathrm{ds}^2}{2Q^2A_0^2}}\label{eq:NdF}
\end{align}
Here, $N_{\Delta f}(f_m)$ is the frequency noise spectrum for the modulation frequency $f_m$, $n_\mathrm{ds}=V_\mathrm{noise}/S_V$ is the noise-equivalent displacement, and $A_0$ is the zero-peak amplitude of the sensor deflection. The second term in the root (detector noise) is a high-frequency component proportional to the square of the modulation frequency $f_m^2$ and does not depend on $Q$. In contrast, the first term (thermal noise) and the third term (oscillator noise) represent the low-frequency components not proportional to $f_m$, which is proportional to $Q^{-1}$ and $Q^{-2}$, respectively. 

\begin{figure}
       \centering
       \includegraphics[width=0.4\columnwidth,page=14]{Figures-clean-crop.pdf}
       \caption{\label{fig:dFspectrum-HT} (a) $\Delta f_\mathrm{LPF}$ and (b) $\Delta f_\mathrm{LO}$ noise spectrum at room temperature and 260 °C using the same setup as in Figure \ref{fig:ThermalExp} with different sensors, along with the theoretical curves from Equation \eqref{eq:NdF}\cite{Kobayashi2009}. The demodulator setting is the same as Figure 4(a). $A_0 (=A_\mathrm{p-p}/2) = \SI{230}{pm}$. At room temperature: $f_0=\SI{20760}{Hz},\ Q=477,\ V_\mathrm{noise}=\SI{3.8}{\micro V\ Hz^{-1/2}}$. At $T_\mathrm{Heater}=\SI{260}{\degreeCelsius}$: $f_0=\SI{20690}{Hz},\ Q=194,\ V_\mathrm{noise}=\SI{4.6}{\micro V\ Hz^{-1/2}},\ T_\mathrm{qPlus}=\SI{179}{\degreeCelsius}$.}
   \end{figure}
Figure \ref{fig:dFspectrum-HT} shows the (a) $\Delta f_\mathrm{LPF}$ and (b) $\Delta f_\mathrm{LO}$ spectra obtained at room temperature and 260 °C using the same setup as in Figure \ref{fig:ThermalExp} with different sensors, along with the theoretical curves from Equation \eqref{eq:NdF}. While there is a slight increase in high-frequency components corresponding to the rise in $V_\mathrm{noise}$, a clear increase in low-frequency noise is observed with the decrease in $Q$. While the temperature limit would vary depending on the required signal-to-noise ratio and the inherent variation in $Q$ between sensors due to individual differences, regardless of temperature, this would be almost the maximum temperature at which the best $\Delta f_\mathrm{LPF}$ signal-to-noise ratio can be obtained. Since the trade-off between $Q$ and $T_\mathrm{Heater}$ is almost unavoidable as long as standard qPlus sensors assembled with heat-resistant epoxy adhesives with the limited $T_g$ are used, atomic-resolution imaging at higher temperatures could require a fundamental modification of the fabrication method for the qPlus sensors, such as the use of inorganic adhesives.

\section{\SI{}{\micro m}-scale calibration and maximum scan range evaluation for Quadpod scanner}
\begin{figure}
 \centering
 \includegraphics[width=0.95\linewidth,page=9]{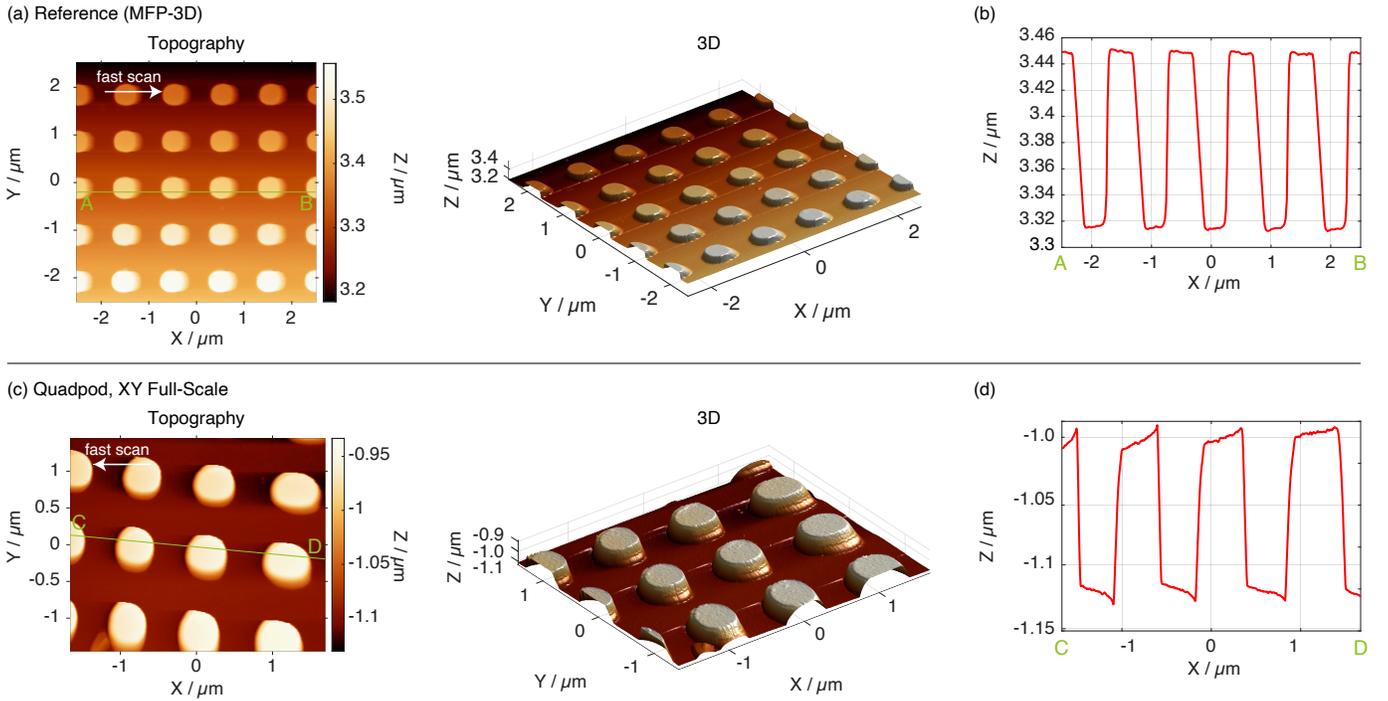}
 \caption{(a) The reference topographic image and its three-dimensional (3D) representation of the used custom-made grid-patterned photomask, obtained in air and at room temperature, using the commercial AFM (MFP-3D-SA, Asylum Research Corp.) with closed-loop scanner and Si microcantilever (SI-DF40, Hitachi High-Tech Corp.). Imaging time: \SI{21}{min\ frame^{-1}}. (b) The line profile along the A-B line in (a). (c) The topographic image and its three-dimensional (3D) representation of the same photomask as (a), obtained in a $\sim 10^1\SI{}{Pa}$ vacuum and at room temperature using the developed Quadpod-scanner-based AFM (Figure 1(c) of the main text) with the Hybrid-loop frequency demodulation (Figure 4(b) of the main text). $f_0=\SI{18629}{Hz}$, $Q=997$, $A_\mathrm{p-p}\sim\SI{233}{pm}$. Imaging time: \SI{25}{min\ frame^{-1}}. (d) The line profile along the C-D line in (c).}
 \label{fig:umscale}
\end{figure}

To estimate the \SI{}{\micro m}-scale calibration and the maximum scan range for AFM experiments, a grid-patterned photomask was imaged using the developed Quadpod-scanner-based AFM. For reference, \textbf{Figure \ref{fig:umscale}(a)} shows the topographic image and its three-dimensional (3D) representation of the used custom-made grid-patterned photomask, which was obtained in air and at room temperature using the commercial AFM (MFP-3D-SA, Asylum Research Corp.) with closed-loop scanner and Si microcantilever (SI-DF40, Hitachi High-Tech Corp.), along with its line profile in \textbf{Figure \ref{fig:umscale}(b)}. The bumps of the Cr mask with a flat top surface are arranged in a grid pattern with a \SI{1}{\micro m} pitch. The height of the bumps is $\sim\SI{130}{nm}$, as confirmed by the line profile shown in Figure \ref{fig:umscale}(b). 

\textbf{Figure \ref{fig:umscale}(c)} shows the topographic image of the same grid-patterned photomask and its 3D representation, obtained in a $\sim 10^1\SI{}{Pa}$ vacuum and at room temperature using the developed Quadpod-scanner-based AFM shown in Figure 1(c) of the main text. The lateral scan range for Figure \ref{fig:umscale}(c) was set to the maximum scan range of the developed setup, for the applied voltage range of $-\SI{40}{V}\leq (V_x, V_y) \leq\SI{40}{V}$ (see the main text for the definitions of the applied voltage). \textbf{Figure \ref{fig:umscale}(d)} shows the line profile of the topographic image of Figure \ref{fig:umscale}(c). Note that, the coordinates in Figure \ref{fig:umscale}(c) and \ref{fig:umscale}(d) along each axis (X and Y for lateral and Z for vertical) have been rescaled so that the grid pitch and bump height are equal to \SI{1}{\micro m} and \SI{130}{nm}, respectively, thereby the \SI{}{\micro m}-scale calibration was determined. \label{page:creep1}The obtained topographic image and its 3D representation in Figure \ref{fig:umscale}(c) clearly show the grid-patterned bump structure, as in the reference image of Figure \ref{fig:umscale}(a). 

    \begin{figure}
        \centering
        \includegraphics[width=0.4\columnwidth,page=13]{Figures-clean-crop.pdf}
        \caption{\label{fig:forcecurve} The frequency shift ($\Delta f_\mathrm{LO}$)-distance curves obtained on the Ga/\ce{GaO_x} interface at room temperature. For approach and retract scan. $f_0=\SI{20751}{Hz}$, $Q=441$, $A_\mathrm{p-p}\sim\SI{133}{pm}$, total acquisition time $=\SI{3.3}{s}$.}
    \end{figure}
Note, as can be seen from the bending of the flat bump top and bottom substrate surfaces in the corresponding line profile in Figure \ref{fig:umscale}(d), the effect of creep cannot be ignored. As with most conventional open-loop scanners, for µm-scale applications beyond high-resolution imaging that do not tolerate creep during µm-scale scans, such as quantitative surface profiling, it is desirable to implement a closed-loop setup using additional capacitive sensors or similar position detectors. On the other hand, for atomic-scale analysis that does not involve large-range scanning, these effects appear negligible in practice. Figure \ref{fig:forcecurve} shows the frequency shift ($\Delta f_\mathrm{LO}$)-distance  curve obtained on a molten Ga/\ce{GaO_x} interface\cite{Amano2023} at room temperature, formed at the contact between a Ga droplet and a bare mica substrate. The periodic oscillations originating from the layered interface structure\cite{Amano2023} are clearly seen, and the approach and retract scans are almost completely overlapping. This demonstrates that the artifacts due to the creep are practically negligible in nm-scale scans. For this reason, open-loop control was employed in this study to avoid noise caused by additional sensors required for closed-loop operation.

From the obtained \SI{}{\micro m}-scale calibration, the maximum scan ranges for each axis were estimated to be \SI{3.4}{\micro m} in the X direction, \SI{2.9}{\micro m} in the Y direction, and \SI{0.92}{\micro m} in the Z direction for the applied voltage range of $-\SI{40}{V}\leq (V_x, V_y) \leq\SI{40}{V}$ and $\SI{40}{V}\leq V_z\leq\SI{110}{V}$, respectively. Note that, although the Quadpod scanner has a symmetrical structure and theoretically equal sensitivity in the X and Y directions, in practice, individual variations in the sensitivity of the piezoelectric actuator ($\pm\SI{15}{\percent}$\cite{PA4FKYW}) introduce differences in the scanner's overall sensitivity. This \SI{}{\micro m}-scale calibration was used as a rough calibration for high-resolution imaging of Ga/\ce{AuGa2} interface in Figure 5 of the main text, before a \SI{}{nm}-scale re-calibration based on the obtained atomic images. 
Here, while the sensitivity in the Z direction should ideally be equal to that of each stacked piezoelectric actuator in the Quadpod scanner from its working principle, the determined maximum scan range is quite smaller than that calculated from the nominal scan range of each actuator\cite{PA4FKYW}, which is $\SI{4}{\micro m}\ (\SI{150}{V;\ typ}) \times \frac{110-40}{150} = \SI{1.86}{\micro m}$. To verify this, we have built the ``clone'' AFM head with the identical design to the original head and performed the same calibration procedure, which yielded the maximum scan range of \SI{2.9}{\micro m} in the X direction, \SI{2.7}{\micro m} in the Y direction, and \SI{0.86}{\micro m} in the Z direction, respectively. Although the maximum scanning range of the clone AFM head for each axis is a little smaller than that of the original head, this difference is within the sensitivity tolerance of $\pm\SI{15}{\percent}$\cite{PA4FKYW} for the used piezoelectric actuators. Therefore, the difference between the estimated maximum scan range of the Quadpod scanner and that calculated from the nominal scan range of each actuator is expected to be inherent to the scanner design. This can be explained by the in-plane distribution of the voltage-to-displacement sensitivity at the actuator's end face\cite{Takahashi1983} or by the suppression of in-plane deformation due to the rigidly fixed actuator's end face\cite{Takahashi1985}. These factors would be addressed by optimizing the Quadpod scanner legs' structure, if the increased manufacturing complexity is acceptable.

\section{SEM-EDS analyses of microstructure and chemical composition of the intermetallic compounds on the Ga/\ce{PtGa_x} interface}
To estimate the chemical composition of the intermetallic phase formed on the Ga/\ce{PtGa_x} interface, \textit{ex-situ} scanning electron microscopy and energy-dispersive X-ray spectroscopy (SEM-EDS) investigations were performed. The experimental procedure was as follows. A Ga droplet was put onto the Pt ribon (\SI{0.1}{mm} thick, \SI{99.95}{\percent}; Nilaco Corp.) cut into $\sim\SI{3}{mm}$ squares, and then heated for \SI{140}{h} in ambient air, at reaction temperature of 220, 100, and \SI{30}{\degreeCelsius} each. After cooling to room temperature, excess liquid Ga was blown off using the \ce{N2} spray gun\cite{Fujita2017} as far as possible. Then, the fabricated specimen was investigated by optical microscope and SEM-EDS (JSM-6500F and JED-2300F; JEOL Ltd.).

\textbf{Figure \ref{fig:EDStop}} shows the optical microscope (OM), SEM secondary electron (SE), and SEM backscattered electron (BSE) compositional images, for each reaction temperature. At a reaction temperature of \SI{220}{\degreeCelsius} (Figure \ref{fig:EDStop}(a)), most of the Ga was consumed in the reaction, and intermetallic compound grains with distinct faceted surfaces were exposed on the specimen's top surface, as shown in the optical microscope image and SE image. The BSE compositional image in Figure \ref{fig:EDStop}(a) showed no clear contrast, except for the surface topography, suggesting that the formed intermetallic phases have a uniform composition. The EDS analysis was also performed for randomly picked grains, highlighted by boxes in the BSE image, which estimated the average chemical composition to be $a_\mathrm{Ga}/(a_\mathrm{Ga}+a_\mathrm{Pt})=92.0\pm\SI{0.8}{\percent}$ (\SI{95}{\percent} confidence interval; $a_\mathrm{M}=$ atomic fractions of $\mathrm{M}=(\mathrm{Ga},\ \mathrm{Pt})$). 
\begin{figure}
 \centering
 \includegraphics[width=0.95\linewidth,page=10]{Figures-clean-crop.pdf}
 \caption{The optical microscope (OM), SEM secondary electron (SE), and SEM backscattered electron (BSE) compositional images of Ga/\ce{PtGa_x} specimen after blow-off procedure (refer to the text). Reaction time $=\SI{150}{h}$, reaction temperature $=$ (a) \SI{220}{\degreeCelsius}, (b) \SI{100}{\degreeCelsius}, (c) \SI{30}{\degreeCelsius}. SEM acceleration voltage $=\SI{5}{kV}$. The boxes in the BSE image indicate the region for EDS analysis (Number of boxes: (a) 8, (b) 2, (c) 4).}
 \label{fig:EDStop}
\end{figure}

At reaction temperatures of \SI{100}{\degreeCelsius} (Figure \ref{fig:EDStop}(b)) and \SI{30}{\degreeCelsius} (Figure \ref{fig:EDStop}(c)), the reaction proceeded significantly slower than at \SI{220}{\degreeCelsius}, and most of the Ga remained as a liquid phase and blown off using \ce{N2} gun before the investigations. Although most of the excess Ga was removed, as shown in the optical (OM) image, the remaining liquid Ga still covers the solid surface, as shown in the SE images. Still, some relatively large grains were observed in the SE and BSE compositional image, for which the chemical composition was estimated by EDS. The average compositions for randomly picked grains, highlighted by boxes in the BSE images, was estimated to be $a_\mathrm{Ga}/(a_\mathrm{Ga}+a_\mathrm{Pt})=87.9\pm\SI{2.2}{\percent}$ for the reaction temperature of \SI{100}{\degreeCelsius} (Figure \ref{fig:EDStop}(b)) and $92.9\pm\SI{0.8}{\percent}$ for \SI{30}{\degreeCelsius} (Figure \ref{fig:EDStop}(c)).

\begin{figure}
 \centering
 \includegraphics[width=0.95\linewidth,page=11]{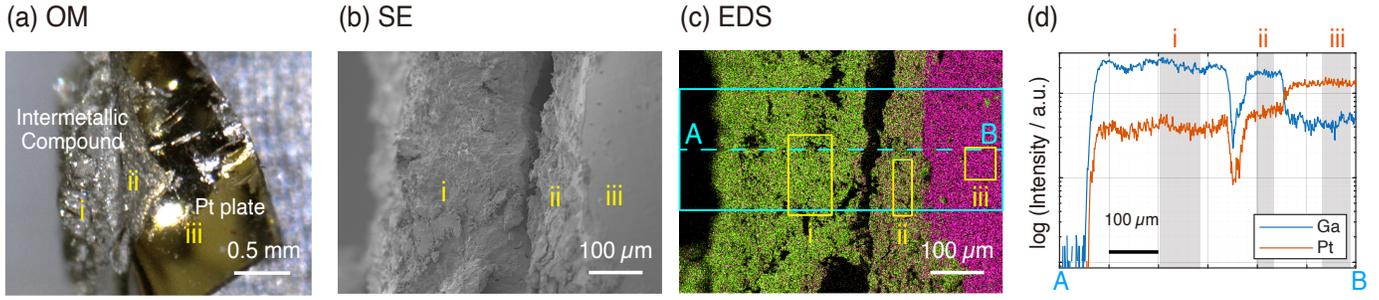}
 \caption{(a, b) (a) The optical microscope (OM) and (b) SEM secondary electron (SE) image obtained on the peeled-off specimen prepared at the reaction temperature of \SI{220}{\degreeCelsius}. (c) The SEM-EDS compositional map of (b). Green and magenta pixels represent the Ga K and Pt M emissions, respectively. (d) The line profiles for the EDS map along the A-B dotted line in (c) (arbitrary unit; a.u.). The blue box in (c) indicates the averaging area for the line profile. The gray area i$-$iii in (d) represents the corresponding area boxed in (c) after orthogonal projection onto the A-B line, respectively. SEM acceleration voltage $=\SI{5}{kV}$.}
 \label{fig:EDScut}
\end{figure}

For all reaction temperatures, the closest match in the widely accepted phase diagram\cite{Li2006,Okamoto2007} is \ce{PtGa6}\cite{Bhan1960}, which agrees with the previous studies\cite{Zhang2017,Yazdanapanah2008}. While the composition estimated by EDS was shifted toward the Ga-rich side compared to the stoichiometry of \ce{PtGa6}, this is explained by residual liquid Ga remaining on the grain surfaces and the gaps between the grains.
In addition, to roughly estimate the elemental distribution along the depth direction, we performed a fracture surface analysis using a peeled-off specimen, which was prepared in the same procedure as in Figure \ref{fig:EDStop}(a) at the reaction temperature of \SI{220}{\degreeCelsius}, then the Pt ribbons was mechanically peeled off from the formed bulk intermetallic compounds. \textbf{Figure \ref{fig:EDScut}(a)} illustrates the optical microscope (OM) images of the peeled-off specimen.
\textbf{Figure \ref{fig:EDScut}(b)} and \textbf{\ref{fig:EDScut}(c)} show the SEM SE image and the corresponding EDS compositional map for the peeled-off specimen of Figure \ref{fig:EDScut}(a), respectively.
Here, the green and magenta pixels in the EDS map in Figure \ref{fig:EDScut}(c) represent the Ga K and Pt M emissions, respectively. 
In addition, the line profile for the EDS map along the A-B line in Figure \ref{fig:EDScut}(c) is shown in \textbf{Figure \ref{fig:EDScut}(d)}.

In the optical and SE images, the field of view is divided into three regions: (i) the bulk intermetallic alloy phase, (ii) the intermetallic phase remaining on the Pt ribbon side, and (iii) bulk Pt, which is also annotated in all the images and the line profile of Figure \ref{fig:EDScut}. The EDS map in Figure \ref{fig:EDScut}(c) shows that the composition is uniform within each of these regions, as evidenced by the corresponding line profile in Figure \ref{fig:EDScut}(d). The average composition for each region, annotated by the yellow boxes in Figure \ref{fig:EDScut}(c), was estimated to be (i) $a_\mathrm{Ga}/(a_\mathrm{Ga}+a_\mathrm{Pt})=\SI{87.1}{\percent}$, (ii) $\SI{75.1}{\percent}$, and (iii) below detection limit (\textit{i.e.} unalloyed Pt), respectively. The closest match for the region (i) and (ii) in the phase diagram\cite{Li2006,Okamoto2007} is \ce{PtGa6} and \ce{Pt3Ga7}, respectively, with some shift toward the Ga-rich side from the stoichiometry as in the top-view investigations in Figure \ref{fig:EDStop}(a). These results indicate that multiple intermetallic compound phases including \ce{PtGa6} and \ce{Pt3Ga7} are formed, while mainly \ce{PtGa6} is exposed on the liquid/solid interface, as confirmed in the top-view analysis in Figure \ref{fig:EDStop} at each temperature.

\bibliographystyle{MSP}